\documentclass[11pt,a4paper]{article}
\usepackage[a4paper, margin=2cm]{geometry}
\usepackage{authblk}
\usepackage{bbm}
\usepackage[utf8]{inputenc}
\usepackage[T1]{fontenc}
\usepackage{stackengine}
\usepackage{amsmath}
\usepackage{amsfonts}
\usepackage{amssymb}
\usepackage{mathtools}
\usepackage{array}
\usepackage{breqn}

\usepackage{floatrow}
\usepackage{graphicx}
\usepackage{tabularx}
\usepackage{multirow}
\usepackage{wrapfig}
\usepackage{titling}
\usepackage{rotating}
\usepackage{float}
\usepackage{color}	
\usepackage[font=footnotesize,labelfont=bf]{caption}
\usepackage{subcaption}
\usepackage{xr}
\usepackage{hyperref}	
\usepackage{todonotes}
\usepackage{rotating}
\usepackage{stackengine}
\usepackage{algorithm}
\usepackage{algpseudocode}
\usepackage{makecell}
\usepackage{afterpage}
\usepackage[numbers,sort]{natbib}

\graphicspath{{./graphics/}}
\usepackage[normalem]{ulem}					
\newcommand{\TWEditMayEighth}[1]{{\color{black} #1}}
\newcommand{\TWEditMayTwentyEighth}[1]{{\color{black} #1}}
\newcommand{\TLText}[1]{{\color{black} #1}}

\def\tinytext#1{\text{\tiny #1}}
\newcommand{\muattrLPSLPS}[0]{\mu_{{\tiny \text{attr}}}^{{\tiny {\text{LPS-LPS}}}}}
\newcommand{\muattrOMPLPS}[0]{\mu_{{\tiny \text{attr}}}^{{\tiny {\text{OMP-LPS}}}}}
\newcommand{\muattrOMPOMP}[0]{\mu_{{\tiny \text{attr}}}^{{\tiny {\text{OMP-OMP}}}}}

\newcommand{\Rsensins}{R^{\tinytext{sens}}}
\newcommand{\diffLPS}{D_{\tinytext{LPS}}}

\title{An agent-based model of outer membrane biogenesis in Gram-negative bacteria}
\author{Thomas Williams$^{1,2}$, James M. Osborne$^{1,2}$, Kwok Jian Goh$^{3,4}$, Trevor Lithgow$^{3,4}$, \\Jennifer Flegg$^{1,2}$}
\date{{\footnotesize$^1$ARC Centre of Excellence for the Mathematical Analysis of Cellular Systems, University of Melbourne, Australia. $^2$School of Mathematics and Statistics, University of Melbourne, Australia. $^3$ARC Centre of Excellence for the Mathematical Analysis of Cellular Systems, Monash University, Australia. $^4$Infection Program, Biomedicine Discovery Institute and Department of Microbiology, Monash University, Australia.}}

\begin{document}
	
	\maketitle
	
	\begin{abstract}
			
			\TLText{The outer membrane is the \TWEditMayEighth{interface through which Gram-negative bacteria --- a broad classification of organisms including \textit{Escherichia coli} and a number of deadly pathogens} --- interact with the environment. Two decades of work on the process of outer membrane biogenesis have led to the discovery of the components that mediate this process, and the \TWEditMayTwentyEighth{characterisation of} structure and function of these component parts of the \TWEditMayTwentyEighth{bacterial cell} machinery.} However, neither current experimental methods, nor conventional molecular dynamics (MD) \TWEditMayTwentyEighth{simulation} approaches are capable of investigating this membrane machinery on the time scale of the cell division cycle. This leaves crucial questions unanswered, such as how this lipid-poor, largely static environment is organised to permit ongoing membrane growth. Here, we introduce a semi-quantitative agent-based model to explore the molecular-scale dynamics of Gram-negative outer membrane as it grows. Model simulations across a broad region of parameter space suggest that protein incorporation into the membrane by the $\beta$-barrel assembly machinery (BAM complex) is a process which is prone to stalling, and may take place only in short bursts. We also find suggestions that BAM complexes work collaboratively with each other, and with the lipopolysaccharide-inserting Lpt complex when in close proximity. The agent-based framework we introduce provides a means to assess and generate hypotheses on outer membrane biogenesis on previously inaccessible time scales.
	\end{abstract}

	\section*{Introduction}
	
	Gram-negative bacteria are distinguished from their Gram-positive counterparts by a characteristic double membrane. The exterior-facing outer membrane (OM) is a proteinaceous structure which mediates the cell's uptake of nutrients, and acts as the cell's first line of defence against threats, including antibiotics \cite{lithgow_et_al_surveying, silhavy_et_al_bacterial}. The OM is majority protein by mass (and, potentially, by surface area \cite{muhlradt_et_al_lateral_mobility}), solubilised in a relatively small quantity of lipid \cite{horne_et_al_lipid_bilayer}. These lipids are arranged asymmetrically, with phospholipids in the inner leaflet and lipopolysaccharides (LPS) in the outer leaflet \cite{horne_et_al_lipid_bilayer}. By some estimates, there are as few as 2--4 LPS molecules and 4--10 phospholipid moleculues per outer membrane protein \cite{lessen_et_al_building}. The integral membrane proteins studded in the OM have a so-called $\beta$-barrel structure, and have been observed under atomic force microscopy to pack in dense, \TWEditMayTwentyEighth{largely} static arrays \cite{benn_et_al_OmpA_order}.
	
	\TWEditMayEighth{The periplasmic space between the outer and inner membranes is devoid of \TWEditMayTwentyEighth{ribosomes and lipid-synthesising enzymes}, thus the OM is constructed from components that are synthesised in the interior of the cell and transported to the membrane by specialised molecular complexes \cite{silhavy_et_al_bacterial}.} \TLText{LPS is transported from the inner membrane, across the periplasmic space and to the OM by the Lpt machinery. The outer membrane part of this lipid transport pathway is \TWEditMayEighth{the $\beta$-barrel protein} LptD \TWEditMayEighth{(paired with the lipoprotein LptE), which} functions to flip LPS monomers into the outer leaflet of the OM \cite{dong_et_al_structural, botos_et_al_structural}. Early work by microscopy showed that LPS insertion creates localised patches \cite{muhlradt_et_al_OM_salmonella, muhlradt_et_al_lateral_mobility}, \TWEditMayEighth{possibly} representing the location of LptDE, and leading to the understanding that membrane growth could be the product of discrete events in which many molecules are inserted in bursts of activity \cite{ursell_et_al_surface_protein}. Recent evaluation of \textit{E. coli} cell surfaces by atomic force microscopy showed these lipid patches in unprecedented detail \cite{benn_et_al_OmpA_order, lithgow_et_al_surveying}.}
	
	\TLText{Outer membrane proteins (OMPs), including LptD and \TWEditMayEighth{others, such as} the highly abundant porin OmpA, \TWEditMayEighth{undergo a similarly complex journey to the OM. OMP polypeptides} are translated \TWEditMayEighth{--- but not folded ---} in the cytoplasm, \TWEditMayEighth{are} translocated through the inner membrane and periplasm to the OM, and there are folded into a $\beta$-barrel \TWEditMayEighth{\TWEditMayTwentyEighth{during incorporation} into the OM} by the $\beta$-barrel assembly machinery (BAM complex) \cite{silhavy_et_al_bacterial, wu_et_al_plasticity, shen_et_al_structural_basis}. The BAM complex consists of the essential core subunit BamA, itself a $\beta$-barrel \cite{noinaj_et_al_structural_insight, zahn_BamA_PDB}, and a set of peripheral lipoproteins that are lineage specific \cite{webb_et_al_evolution, anwari_et_al_evolution_lipoprotein}. Early studies showed that OMPs appear at the cell surface in bursts \TWEditMayEighth{(as do LPS)} and that the site of \TWEditMayEighth{insertion} is distributed randomly across the bacterial surface \cite{ursell_et_al_surface_protein}. \TWEditMayEighth{Likewise,} super-resolution imaging of the \textit{E. coli} cell surface showed that the BAM complex is distributed across the bacterial surface and that its lateral movement is very slow \cite{gunasinghe_et_al_WD40_protein, lee_upton_et_al_slow_diffusion_BAM}. \TWEditMayEighth{However, the spatial distribution of BAM complexes is not entirely random:} imaging as well as \textit{in situ} cross-linking showed that while some BAM complexes may be discrete units mediating $\beta$-barrel assembly, many BAM complexes are arranged in assembly precincts where each BamA subunit is within chemical cross-linking distance of several others \cite{gunasinghe_et_al_WD40_protein}.} \TWEditMayEighth{This arrangement has been proposed to aid in the assembly of multimeric OMPs, such as the abundant trimeric proteins OmpC and OmpF in \textit{E. coli} \cite{gunasinghe_et_al_WD40_protein}.}
	
	\TWEditMayEighth{Molecular dynamics (MD) simulation has further improved our understanding of outer membrane biogenesis at a molecular and sub-molecular scale, and of OMP assembly in particular \cite{khalid_et_al_atomistic_and_coarse, im_and_khalid_molecular_simulations, webby_et_al_lipids, gutishvili_et_al_seeing_is_believing}.} \TLText{MD simulations of BamA in membrane environments illustrated and modelled a zippering of the N- and C-terminal $\beta$-strands of BamA \cite{iadanza_et_al_distortion, liu_and_gumbart_membrane_thinning}. This \TWEditMayEighth{so-called ``lateral gate''} serves at least two functions: thinning and destabilising the lipids in the immediate vicinity of BamA to assist the energetics of the substrate polypeptide entering the membrane, and opening sufficiently to provide a template for each $\beta$-strand of the substrate to fold as the circularisation of the $\beta$-barrel ensues \cite{kuo_et_al_modelling_intermediates, wu_et_al_plasticity, iadanza_et_al_distortion, shen_et_al_structural_basis, doyle_et_al_cryoEM}.}
	
	While MD has proved a hugely successful approach for studying the structural and physical features of the OM, its steep computational cost limits the temporal scale it is capable of modelling. Despite efficient codes and hardware acceleration, all-atom MD simulations of the OM are typically limited to a time scale on the order of 100ns--1$\mu$s \cite{chavent_et_al_MD_simulations}. Even coarse-grained MD, in which groups of atoms are bundled into submolecular ``beads'', is rarely used to simulate membrane systems with more than 100 proteins for longer than 10$\mu$s \cite{chavent_et_al_MD_simulations, im_and_khalid_molecular_simulations}. This strict time scale limitation makes MD an unsuitable tool to probe the dynamics of the OM on the seconds-to-minutes time scale of membrane growth.
	
	Within the time scale \TWEditMayTwentyEighth{of the cell division cycle} --- which is too long for MD to describe, and too fine to access experimentally --- lie key questions. For example, what happens to OMPs once they leave the BAM? How does the OM protein array emerge, and how is it maintained? Perhaps most interestingly: given the apparent necessity of LPS for protein assembly in the OM \cite{horne_et_al_lipid_bilayer, doyle_et_al_cryoEM, peterson_et_al_conserved}, how is a supply kept available to the BAM complexes when lipids are so scarce and lateral diffusion is so minimal?
	
	Off-lattice agent-based modelling, with entire molecules each represented by a single agent, is a natural framework to address such a problem and has a far lower computational cost per model time compared to MD. Approximating the plane of the OM as 2D --- and, consequently, agents as circles --- offers a further reduction in computational complexity. Some attempts have so far been made in this direction. Chavent \textit{et al.} \cite{chavent_et_al_nanoscale} and Dunton \cite{dunton_thesis} both developed 2D agent-based models for systems of OMPs of a single species (BtuB and NanC, respectively) in phospholipid bilayers, which permitted insights into the formation of protein clusters and the influence of protein interaction on molecular diffusion. Rassam \textit{et al.} also studied diffusion and crowding dynamics in a model which tracked BtuB, OmpF and OmpA \cite{rassam_et_al_supramolecular}. None of these models considered membrane growth or explicitly included lipids in the system \cite{chavent_et_al_nanoscale, dunton_thesis}. The latter assumption is reasonable for \textit{in vitro} systems comprised of small, abundant, and highly motile phospholipids, but do not apply to the \textit{in vivo} OM, whose outer leaflet lipids consist only of LPS \cite{horne_et_al_lipid_bilayer}. LPS molecules are of a similar length scale and found in similar quantities to proteins in the OM \cite{horne_et_al_lipid_bilayer, lessen_et_al_building}, thus it is reasonable to expect that the dynamics of individual LPS molecules will have a meaningful impact on the system.
	
	In this work, we introduce an agent-based model of a growing Gram-negative OM at molecular (but not submolecular) resolution, in which OMPs \textit{and} LPS are treated as discrete agents. This level of granularity in the model makes it computationally feasible to probe the dynamics of the OM as it grows, while capturing essential aspects of the underlying biophysics. With this framework, we explore the emergent protein structure of the OM, and the consequences of BAM-mediated protein insertion.

	\section*{Methods}
	
	We developed a stochastic, spatially-explicit, agent-based model of simplified molecular dynamics in the outer aspect of the Gram-negative outer membrane. We outline key features of the model here, and provide full details in Supplementary Section S2. Our model is implemented in Julia, including support for CUDA GPU acceleration, and is freely available from GitHub\footnote{\url{https://github.com/thomaswilliams23/MembraneABMSim/tree/paper_1}}.
	
	\subsection*{Overview}
	The model tracks a patch of membrane occupied by both protein (OMP) and lipid (LPS) molecules. We model three OMP species --- OmpA, BamA, and LptD --- and a single type of LPS. Each molecule is represented by a circular agent with area equal to that molecule's cross-sectional area in the plane of the membrane, \TWEditMayTwentyEighth{calculated from} molecular structures \TWEditMayTwentyEighth{available from the Protein Data Bank} \cite{pautsch_OmpA_PDB, lessen_et_al_building, dong_et_al_structural, albrecht_et_al_BamA_structure, lebrun_et_al_structural}. Exact agent sizes are listed in Supplementary Section 3. \TWEditMayEighth{Agents inhabit} a square spatial domain whose size we allow to vary over time, and we assume periodicity in the horizontal and vertical directions to avoid introducing boundary effects. \TWEditMayEighth{Note that throughout this manuscript, and in the supporting materials, we use the term ``domain'' to refer to the region of physical space occupied by the membrane, in which agents are allowed to reside (that is, the mathematical sense of the term), and \textit{not} a structural unit of a protein (the molecular biology sense of the term).}
	
	\subsection*{Intermolecular forces}
\begin{figure}[h!]
		\centering
		\includegraphics[width=\textwidth, page=1, trim={2cm, 11.2cm, 2cm, 2cm}, clip]{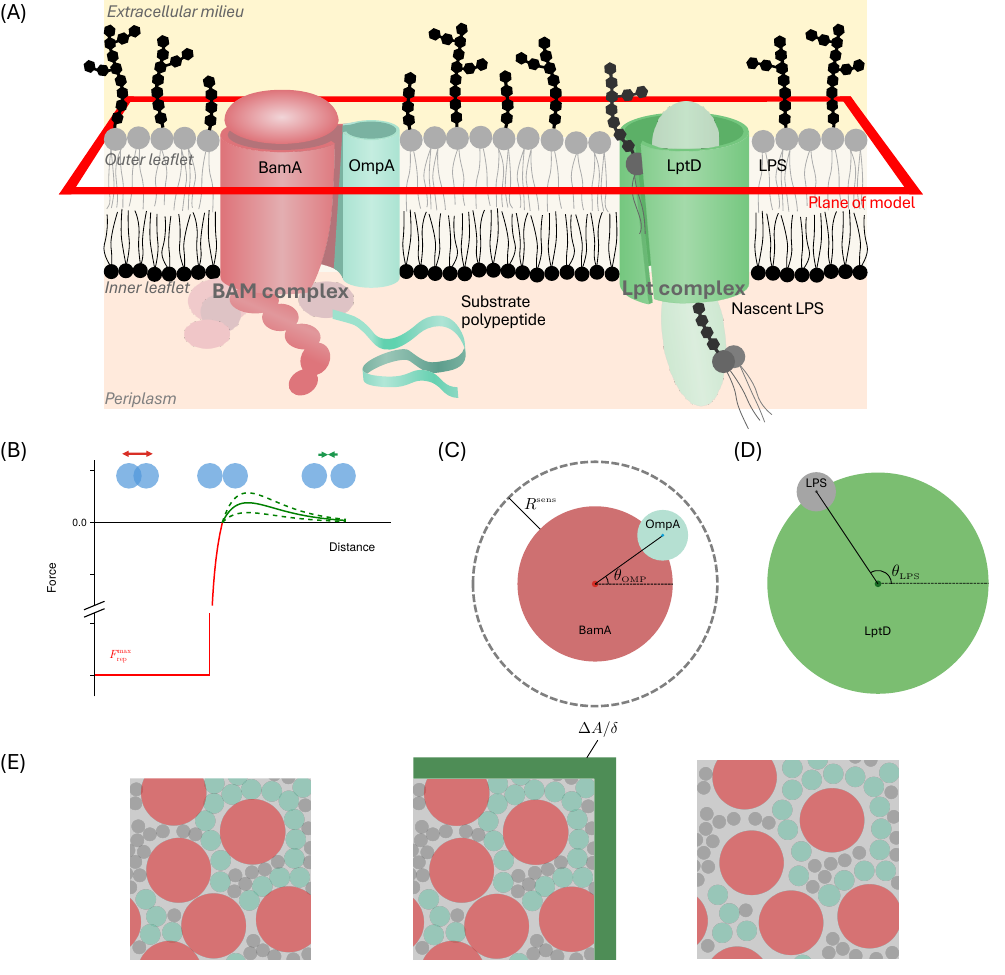}
		\caption{\textbf{Key aspects of the biology and the model.} (A) Biology of the growing Gram-negative OM. The $\beta$-barrel assembly machinery (BAM) complex (left) is supplied with unfolded polypeptides from the periplasm, which it folds into new $\beta$-barrel proteins in the OM (such as OmpA, pictured). Likewise, the Lipopolysaccharise translocase (Lpt, right) supplies the OM with lipopolysaccharides (LPS). (B) General shape of the force curve between a pair of agents in the model as a function of the distance between their centres. Nearby agents experience a weak attraction force (green curves, we show curves for three values of the attraction force coefficient $\mu_{\tinytext{attr}}^{(\cdot\cdot)}$); overlapping agents experience a strong repulsion up to some maximum quantity $F_{\tinytext{rep}}^{\tinytext{max}}$ (red curve). When the distance between two agents is exactly the sum of their radii, no force is experienced. The attraction forces for different agent type combinations (LPS-LPS, OMP-OMP, OMP-LPS) are characterised by different parameters. (C) OMP insertion mechanism in the model. BamA (red) can insert a new OMP (pale green) only if some part of an LPS lies within the sensing radius $\Rsensins$ of the BamA. The new OMP is inserted at some random angle $\theta_{\tinytext{OMP}}$ from the horizontal. (D) LPS insertion mechanism in the model. LptD inserts LPS via a similar mechanism to BAM (along a random angle $\theta_{\tinytext{LPS}}$ to the horizontal), but without the requirement for nearby LPS. (E) Membrane resizing in the model. The \TWEditMayEighth{computational} domain is periodic in the horizontal and vertical directions. If, during some time step, $\Delta A$ of agent area is added to the system, we add $\Delta A/\delta$ of area to the \TWEditMayEighth{computational} domain, where $\delta$ is a packing density parameter. Each agent's position is then dilated such that its $x$ and $y$ coordinates proportional to the dimensions of the domain are preserved. Growth \TWEditMayEighth{of the computational domain} is blocked if a sufficiently large unoccupied region, i.e. a hole, is present.}
		\label{fig:model_demo}
	\end{figure}

	The motion of agents is mainly determined by the attraction-repulsion forces between pairs of agents. Nearby, non-overlapping agents experience a weak attraction force, and agent overlap --- although permitted under the model --- is subject to a strong repulsive force. The mathematical formulation of the attraction-repulsion force between any pair of agents is given by Supplementary Equation (S13) in Supplementary Section S2.2.9. Associated parameter values are listed in Supplementary Section S3. Of particular importance to the present study is the coefficient of the attraction force, $\mu_{\tinytext{attr}}^{(ij)}$, which \TWEditMayEighth{governs the affinity} between a pair of agents $i$ and $j$. In this work we assume $\mu_{\tinytext{attr}}^{(ij)}$ varies depending on the type of agents involved. We consider three cases: $\mu_{\tinytext{attr}}^{(ij)}=\muattrLPSLPS$ if both $i$ and $j$ are LPS, $\mu_{\tinytext{attr}}^{(ij)}=\muattrOMPLPS$ if one of $i$ and $j$ is an LPS and the other is an OMP, and $\mu_{\tinytext{attr}}^{(ij)}=\muattrOMPOMP$ if both $i$ and $j$ are OMPs. For simplicity, we assume all other parameters governing the strength of forces do not depend on the agent types involved. The general shape of the attraction-repulsion curve is illustrated in Figure~\ref{fig:model_demo}(B), and individual curves for the different agent type combinations (with OmpA as a representative OMP) are plotted in Supplementary Figure S7 (in Supplementary Section S2).

	\subsection*{Insertion of new molecules into the outer membrane}
	\TLText{Throughout this work, for simplicity, we consider the insertion of a single type of substrate polypeptide, namely, the OMP OmpA \cite{benn_et_al_OmpA_order}}. We model the insertion of new OMPs into the membrane as follows. Each BamA agent is assumed to be part of an assembled BAM complex, which is modelled as being in one of three states: \textit{free}, \textit{embedding}, or \textit{stalled}. We assume that a \textit{free} BAM complex randomly becomes bound to an available substrate polypeptide. This rate is independent of the position of the complex, \TLText{since monitoring of OMP insertion in vivo as well as monitoring of BAM activity in vivo has shown that the embedding state occurs randomly across the surface of an E. coli cell \cite{ursell_et_al_surface_protein, gunasinghe_et_al_WD40_protein}}.  Once bound to a substrate polypeptide, the BAM complex begins \textit{embedding} \TWEditMayEighth{the polypeptide as} a new OMP only if there is an LPS within some radius $\Rsensins$ of the BamA agent --- this reflects suggestions from the literature that insertion of new OMPs by BAM is reliant on the presence of local LPS \cite{lithgow_et_al_surveying, horne_et_al_lipid_bilayer}. If sufficiently nearby LPS exists, the BAM complex will then begin inserting a new nascent OMP into the \TWEditMayEighth{computational} domain \TWEditMayEighth{from a random point on its circumference} (at angle $\theta_{\tinytext{OMP}}$ to the horizontal) before returning to the \textit{free} state. If not, the BAM complex becomes \textit{stalled} until an LPS comes within $\Rsensins$ of the BamA agent, allowing it to switch to the \textit{embedding} state. \TWEditMayEighth{Under this process, which takes place over some insertion period, nascent OMPs} \TLText{emerge gradually as a \TWEditMayTwentyEighth{crescent} shape that will ultimately form a \TWEditMayTwentyEighth{complete} circle,} \TWEditMayEighth{in agreement with experimental and MD analyses of BAM-mediated OMP insertion \cite{shen_et_al_structural_basis, doyle_et_al_cryoEM, wu_et_al_plasticity}. Our model of} the OMP insertion process is illustrated in Figure~\ref{fig:model_demo}(C). For full details, see Supplementary Section S2.2.4
	
	In some simulations, we also consider LPS insertion by the Lpt \TWEditMayEighth{complex} (which \TWEditMayEighth{is} represented in the model by \TWEditMayEighth{the intergral membrane protein} LptD). We model this process in a similar manner to OMP insertion (with insertion \TWEditMayEighth{from a random point on its circumference, at angle} $\theta_{\tinytext{LPS}}$ to the horizontal), but without a requirement for nearby LPS. \TWEditMayEighth{The assumption of LPS emergence alongside LptD is broadly consistent with current biological understanding of the action of the LptD complex \cite{fiorentino_et_al_dynamics_LPS_translocon, botos_et_al_structural}.} We illustrate our description of Lpt dynamics in the model in Figure~\ref{fig:model_demo}(D). See Supplementary Section S2.2.5 for full details.

	\subsection*{Membrane growth}
	\TWEditMayEighth{Ultimately, we seek to relate the growth of a small representative patch of membrane to the process of membrane growth \TWEditMayTwentyEighth{over a period of time on the order of the cell division cycle}.} Since we consider a periodic \TWEditMayEighth{computational} domain and a growing agent population, it is important to dynamically resize the domain to ensure the membrane is closed, but not overcrowded \TWEditMayEighth{(such that overlaps occur)}. We account for this resizing as follows. In each time step of a model simulation, we tally the agent area to be added this time step from all nascent agents in the process of being inserted. Call this quantity $\Delta A$. We then add $\Delta A / \delta$ in area to the \TWEditMayEighth{computational} domain (where $\delta$ is a packing density parameter), and dilate the positions of all agents such that their positions proportional to the dimensions of the domain are preserved. However, to prevent \textit{too much} area being added to the \TWEditMayEighth{computational} domain, we skip this growth process if any sufficiently large unoccupied regions (holes) have formed in the membrane. \TWEditMayEighth{This reflects the fact, that, in reality, no hole can be tolerated in the membrane without compromising its function.} This is \TWEditMayEighth{implemented in the model} by determining whether a circle of radius $R^{\tinytext{hole}}$ can be placed in the domain without intersecting any agent (here, we assume $R^{\tinytext{hole}}$ to be roughly the same as the radius of an OmpA, \TWEditMayEighth{which is around 14\r{A}} at the surface of the OM \cite{pautsch_OmpA_PDB}). The domain resizing process is illustrated in Figure~\ref{fig:model_demo}(E). For full details see Supplementary Section S2.2.7. The simulation initialisation process, similarly, ensures that the initial size of the \TWEditMayEighth{computational} domain is well-calibrated to the packing density of the agents. This process is detailed in Supplementary Sections S2.2.1--S2.2.3.

	\subsection*{Parameters and units}
	\TWEditMayEighth{Some} model parameters (such as molecular sizes) \TWEditMayEighth{were available} directly from the literature \TWEditMayEighth{and public databases}. However, many of these parameters are not well characterised by available data. In particular, the rates at which OMPs and LPS are added into the Gram-negative OM \textit{in vivo} are not known, and the interaction forces between outer membrane molecular species have not been well characterised at the $\mathcal{O}(1-100\text{\r{A}})$, seconds-to-minutes, ``mesoscopic'' scale that we model in this work (although \textit{micro}scopic-scale descriptions have been studied in detail in the molecular dynamics literature \cite{khalid_et_al_atomistic_and_coarse, im_and_khalid_molecular_simulations}). We approach this parameter uncertainty by exploring the model dynamics across a wide range of reasonable force parameters, and by selecting (approximately equal) OMP and LPS insertion rates such that the system has sufficient time to relax to an energy minimum between insertion events.
	
	As such, the model as presented in this study can be thought of as being \textit{semi-quantitative}: it is sufficient to predict the potential emergent dynamics of the membrane that arise from its constituent assumptions, but cannot be assumed to predict a precise number of OMPs inserted into the membrane per minute, for example. As a consequence, throughout this study, we have chosen to omit units from plots. For more details on the parameters and units of the model, see Supplementary Section S3.

	\section*{Results}

	\subsection*{OMP insertion dynamics of an isolated BAM complex in abundant LPS}
	
	In order to examine OMP insertion dynamics as captured by the model, we \TWEditMayEighth{first} simulated the activity of a single isolated BAM complex in \TWEditMayEighth{an environment rich with} LPS. To capture the parameter uncertainty in the forces between agents, we tested values for the LPS-LPS attraction force coefficient ($\muattrLPSLPS$) --- \TWEditMayEighth{effectively, testing the LPS-LPS affinity} --- over three orders of magnitude. In each simulation, we initialised the system with a single BamA agent (representing a complete BAM complex) among 200 LPS agents and ran ten simulation replicates up to $t=500$ in arbitrary time units.
	
\begin{figure}[h!]
		\centering
		\includegraphics[width=0.98\textwidth, page=2, trim={2cm, 7.7cm, 2cm, 1.8cm}, clip]{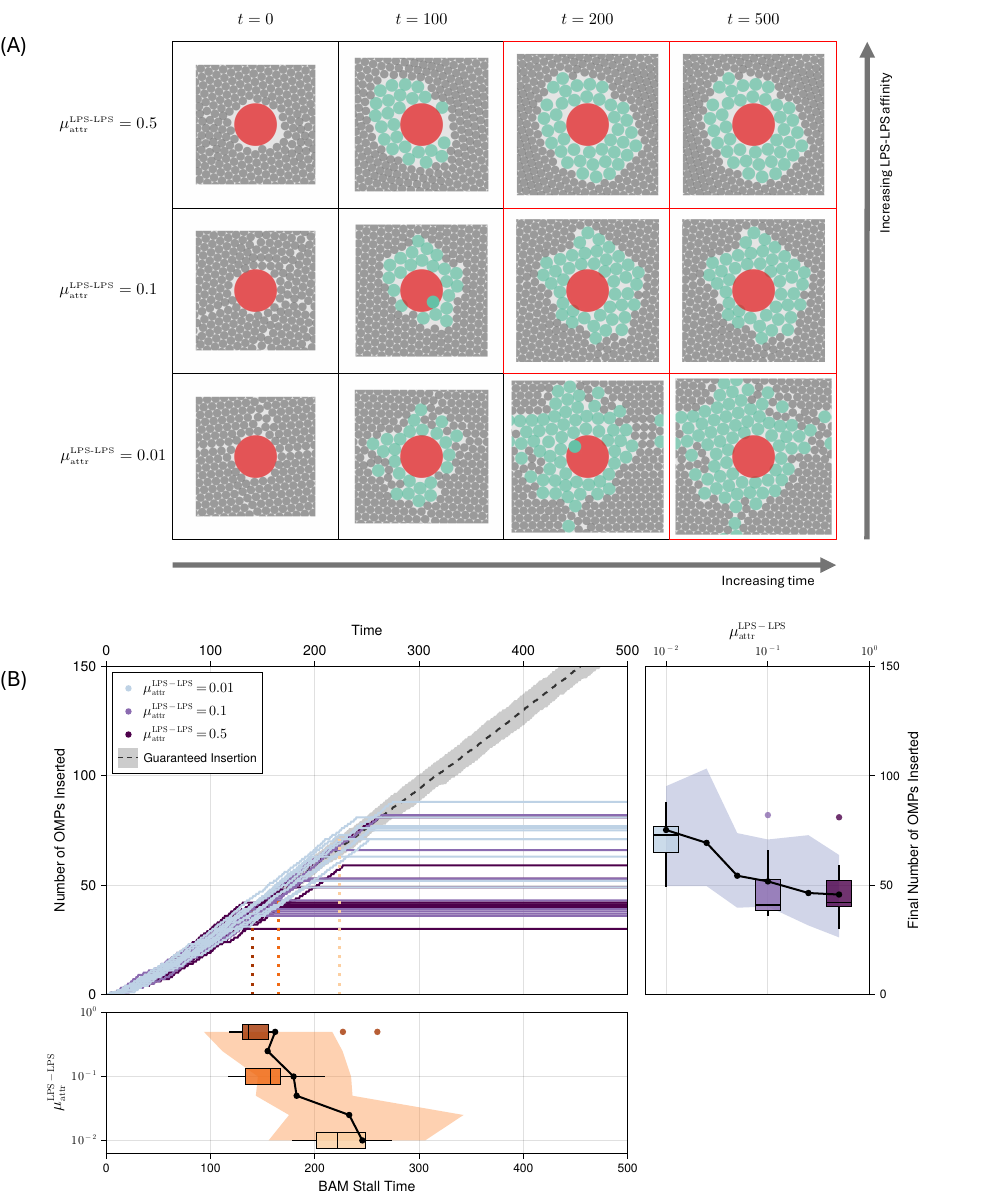}
		\caption{\textbf{Dynamics of an isolated BAM complex in abundant LPS: temporal dynamics and effect of varying LPS-LPS \TWEditMayEighth{affinity}.} (A) Snapshots of a representative simulation over time ($x$-axis) for each of three choices of LPS-LPS attraction force coefficient, $\muattrLPSLPS$ ($y$-axis). LPS in grey, OmpA in pale green, BamA in red. Red grid outlines indicate that the BAM complex no longer has access to LPS within its sensing radius and is now stalled. (B) Top left: number of OMPs inserted over time for each of the choices of $\muattrLPSLPS$ in (A); ten simulation replicates shown for each. For comparison, we also show the mean number of OMPs inserted over time (black dashed line) together with the 0.025--0.975 quantiles (shaded grey region) for ten simulations in which OMP insertion does not depend on LPS proximity. Top right: distribution of final number of OMPs inserted in each of the simulations (as of $t=500$) as a function of $\muattrLPSLPS$; we also show mean values (black line) and the 0.025--0.975 quantiles (shaded region) from data collected from a finer-resolution sweep of $\muattrLPSLPS$ values. Bottom left: time at which the BAM complex stalled due to a lack of nearby LPS; same format as top right plot.}
		\label{fig:LPS-LPS_stickiness_time_series}
	\end{figure}

	The results of these simulations are shown in Figure~\ref{fig:LPS-LPS_stickiness_time_series}. Figure~\ref{fig:LPS-LPS_stickiness_time_series}(A) shows snapshots of the model membrane over time for three choices of LPS-LPS \TWEditMayEighth{affinity}. In each case, the BAM complex is initially highly productive --- inserting dozens of new OMPs into the membrane --- but becomes stalled by the end of the simulation, since the population of OMPs surround the BAM complex, cutting it off from the nearby LPS needed for insertion. This can also be clearly seen from the time series of the number of OMPs inserted in each simulation, which we show in Figure~\ref{fig:LPS-LPS_stickiness_time_series}(B). The plot shows that, regardless of the level of LPS-LPS \TWEditMayEighth{affinity}, OMP insertion in all simulations reach an abrupt cutoff, after which no further OMPs can be added: \TWEditMayEighth{the BAM complex has stalled and does not restart.} As a point of comparison, we show on the same axes the time series for the number of OMPs inserted for a system where OMP insertion is not dependent on LPS proximity. OMP insertion remains steady throughout simulations in this case.
	%
	
\begin{figure}[h!]
		\centering
		\includegraphics[width=\textwidth, page=3, trim={2cm, 22.3cm, 2cm, 2.1cm}, clip]{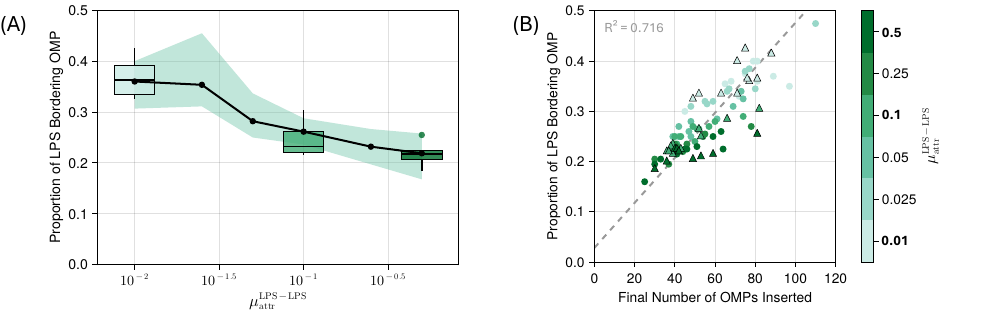}
		\caption{\textbf{Dynamics of an isolated BAM complex in abundant LPS: relationship between BAM productivity and spatial arrangement of LPS.} (A) Proportion of LPS agents bordering an OMP at the end of the simulation ($t=500$) for varying choices of LPS-LPS attraction force coefficient, $\muattrLPSLPS$. Simulations shown in Figure~\ref{fig:LPS-LPS_stickiness_time_series} given as box plots; we also show mean values (black line) and the 0.025--0.975 quantiles (shaded region) from data collected from a fine-resolution sweep of $\muattrLPSLPS$ values (as in Figure~\ref{fig:LPS-LPS_stickiness_time_series}(B)). (B) Final number of OMPs inserted versus final proportion of LPS bordering an OMP. Simulations shown in Figure~\ref{fig:LPS-LPS_stickiness_time_series} given as triangles; data from finer-resolution sweep given as circles. We also show a line of best fit (using the proportion of LPS bordering OMP as a predictor of the final number of OMPs inserted).}
		\label{fig:LPS-LPS_clustering}
	\end{figure}

	While the LPS-LPS \TWEditMayEighth{affinity} does not qualitatively change the long-term outcome of simulations, we did observe a quantitative change in the number of OMPs that could be added to the membrane before the BAM complex became stalled. In the side plots of Figure~\ref{fig:LPS-LPS_stickiness_time_series}(B), we show the distribution of BAM stall times (bottom) and final number of OMPs inserted (right) across different choices for $\muattrLPSLPS$. These plots indicate that the BAM complex remains productive for longer and can insert more OMPs overall when LPS-LPS stickiness is lower. The snapshots of the final system states in Figure~\ref{fig:LPS-LPS_stickiness_time_series}(A) suggest an explanation for this increased productivity. When LPS-LPS \TWEditMayEighth{affinity} is lower, LPS molecules more readily mix into the OMP cluster, thus remaining close to the BAM complex for longer. To quantify this, we analysed the proportion of LPS that were bordering an OMP \TWEditMayTwentyEighth{(that is, were within a short distance, taken to be the LPS radius, of an OMP)} at the end of each simulation. In Figure~\ref{fig:LPS-LPS_clustering}(A) we plot the distribution of this quantity across simulations for different choices of $\muattrLPSLPS$. This plot confirms that LPS border OMPs more frequently at lower LPS-LPS \TWEditMayEighth{affinity}. In Figure~\ref{fig:LPS-LPS_clustering}(B), we show a scatter plot of the proportion of LPS bordering an OMP against the number of OMPs at the end of each simulation, which shows a clear positive relationship between the two quantities.
	%

	We also investigated the effect of changing the two other attraction force coefficients (\TWEditMayEighth{thereby varying} the OMP-OMP and OMP-LPS \TWEditMayEighth{affinities}). We repeated the simulations detailed above while also varying $\muattrOMPOMP$ and $\muattrOMPLPS$ over several orders of magnitude with qualitatively similar results to those presented above. The results of these simulations are shown in Supplementary Figure S1, and show that reducing OMP-OMP \TWEditMayEighth{affinity} or \textit{increasing} OMP-LPS \TWEditMayEighth{affinity} has a similar effect to reducing LPS-LPS \TWEditMayEighth{affinity}, in that the proportion of LPS bordering an OMP is increased and the BAM complex can remain productive for longer. However, we found that the BAM complex was almost guaranteed to stall, regardless of the parameter combination. We extended all simulations out to $t=1000$ time units, by which point all but one simulation replicate had stalled, with the median stall time for each parameter combination under $500$ time units.

	\subsection*{Effect of LPS diffusion on OMP insertion}
	
\begin{figure}[h!]
		\centering
		\includegraphics[width=\textwidth, page=4, trim={2cm, 13.7cm, 2cm, 2.1cm}, clip]{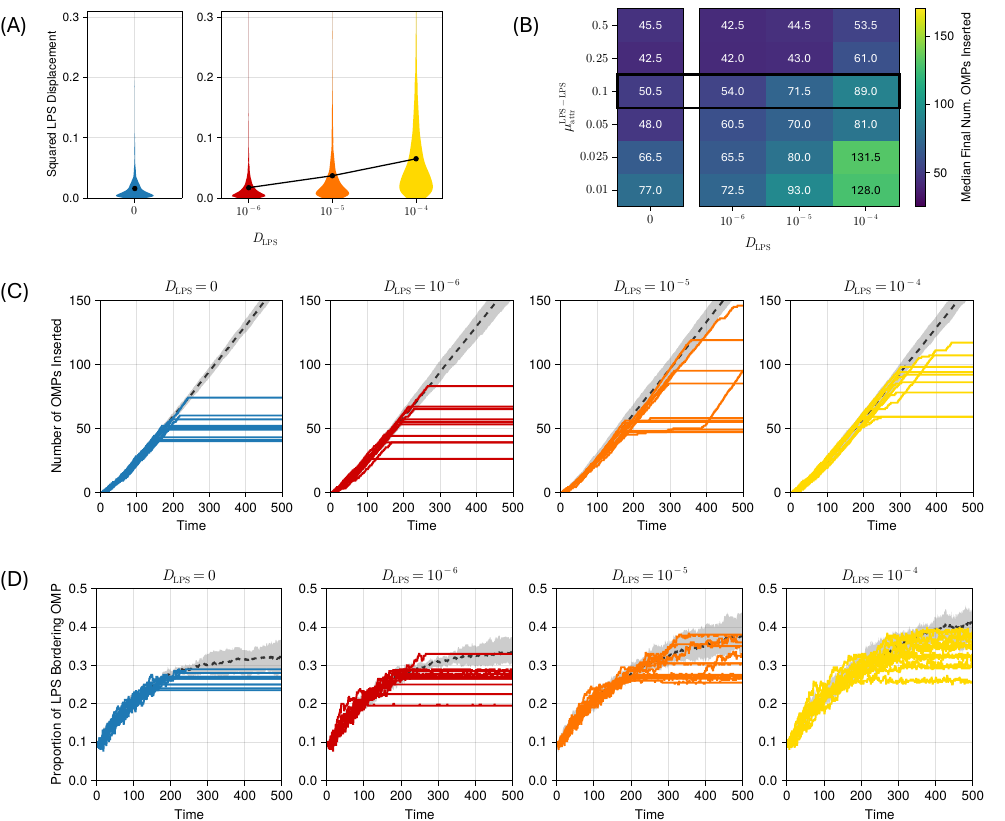}
		\caption{\textbf{Dynamics of an isolated BAM complex in abundant LPS with varying LPS diffusion.} (A) Distribution of squared LPS displacement over the course of a simulation for varying choices of the LPS diffusion coefficient $\diffLPS$, aggregated across ten simulations each. Means indicated in black. (B) Median final number of OMPs inserted (as at $t=500$) across ten simulations for combinations of values of LPS-LPS attraction force coefficient ($\muattrLPSLPS$), and $\diffLPS$. Black cell outlines indicate parameter combinations shown in (A), (C) and (D). (C) Number of OmpAs inserted over time for different choices of $\diffLPS$. We plot trajectories from ten simulations in each case, along with the mean (dashed black line) and 0.025--0.975 quantiles (shaded grey region) for simulations where OMP insertion is not dependent on LPS proximity. (D) Proportion of LPS bordering an OMP over time. Same format as (C).}
		\label{fig:diffusion_sweep}
	\end{figure}

	We explored whether introducing noise --- in the form of explicit agent diffusion --- was sufficient to sustain BAM productivity. We repeated the simulation setup outlined above (a single BamA among 200 LPS) but with the addition of Brownian noise in the motion of the LPS. Since \TWEditMayEighth{measurements in intact bacterial cells show that} OMPs undergo very little lateral diffusion in the OM \cite{ginez_et_al_fluidity_e_coli, lill_et_al_confined_mobility, straatsma_and_soares_characterisation_OprF, lee_upton_et_al_slow_diffusion_BAM}, we did not model any explicit Brownian motion of the OMPs. For full details on the implementation of explicit diffusion in the model, see Supplementary Section S2.2.8. We considered diffusion coefficients $\diffLPS$ for the LPS Brownian motion at three different orders of magnitude, and compared the resulting dynamics from each case with that of the model without explicit diffusion (that is, $\diffLPS=0$). In Figure~\ref{fig:diffusion_sweep}(A) we plot the distribution of squared displacements of the LPS for each choice of $\diffLPS$. These plots indicate that lateral LPS displacement in the membrane occurs even without explicit diffusion in the model (due to membrane repacking caused by OMP insertion), but the displacement observed is substantially increased for the choices of diffusion coefficient tested.
	
	We found that increased LPS diffusion extended the productive period of the BAM complex, but could not prevent it from eventually stalling. In Figure~\ref{fig:diffusion_sweep}(B), we show the median total number of OMPs inserted per simulation across a range of LPS-LPS attraction force coefficients ($\muattrLPSLPS$) and LPS diffusion coefficients. This plot indicates that increased LPS diffusion and decreased LPS-LPS \TWEditMayEighth{affinity} both act to increase the number of OMPs inserted during simulations. However, BAM activity still stalled in all simulations, even for the highest LPS diffusion coefficient tested. This can be seen in the time series plots for the number of inserted OMPs across a range of values for $\diffLPS$ (and a fixed value for $\muattrLPSLPS$), which we show in Figure~\ref{fig:diffusion_sweep}(C). Importantly, there is a clear distinction between the stalling dynamics of the high-diffusion systems and those with low or no explicit diffusion. 
	
	When no explicit source of diffusion is present, or when $\diffLPS$ is small, membrane repacking can only occur from OMP insertion events. In such a system, once the BAM complex is too far from any LPS and becomes stalled, there is no way for the membrane to rearrange to supply LPS to the BAM complex, and the OMP number plateaus (as can be seen in the plots on the left of Figure~\ref{fig:diffusion_sweep}(C)). On the other hand, if $\diffLPS$ is sufficiently large, this can allow membrane repacking to occur even after the BAM complex becomes stalled, which, rarely, can temporarily rescue the complex from its stalled state. This can be seen in some of the trajectories in the plots on the right of Figure~\ref{fig:diffusion_sweep}(C).
	
	We again examined the proportion of LPS bordering OMPs in each of the simulations, this time as time series. These are shown in Figure~\ref{fig:diffusion_sweep}(D). These plots show that the proportion of LPS bordering an OMP plateaus in time. Although the height of the plateau increases with $\diffLPS$, the same general shape was observed for all diffusion coefficients tested. Interestingly, even in simulations where we removed the dependence of BAM activity on nearby LPS (thereby allowing OMP insertion to continue throughout the simulation), the proportion of LPS bordering an OMP still flattened out over the course of simulations. This suggests that while increased LPS diffusion can somewhat increase the extent to which LPS can penetrate the OMP cluster, it is unable to do so quickly enough to allow LPS to remain close to the BAM complex and sustain OMP insertion.

	\subsection*{Dynamics of a system with BAM and Lpt complexes in abundant LPS}	
	
	The simulations we have presented so far have considered a BAM complex in a fixed LPS population. We sought to investigate the effect of also including an LPS source (that is, the Lpt complex) in the simulation, and whether this would be sufficient to maintain BAM activity. We carried out simulations of the model with a single BamA and LptD (the outer membrane component of the Lpt complex) among an initial population of 200 LPS molecules. We again tested values of the LPS-LPS, OMP-OMP and OMP-LPS attraction force coefficients over several orders of magnitude. Snapshots of final system states across three values of $\muattrLPSLPS$ are shown in Figure~\ref{fig:BAM_vs_Lpt}(A). In Figure~\ref{fig:BAM_vs_Lpt}(B) we show the time series of the number of OMPs inserted, extended out to $t=1000$, across a range of $\muattrLPSLPS$ values. Figure~\ref{fig:BAM_vs_Lpt}(B) shows similar dynamics to the high-diffusion simulations above, in that the BAM complex quickly stalls (within around $200$-$400$ time units), but can then be occasionally rescued from the stalled state and resume OMP insertion in short bursts.
	%
	\begin{figure}[h!]
		\centering
		\includegraphics[width=\textwidth, page=5, trim={2cm, 6.2cm, 2cm, 2.1cm}, clip]{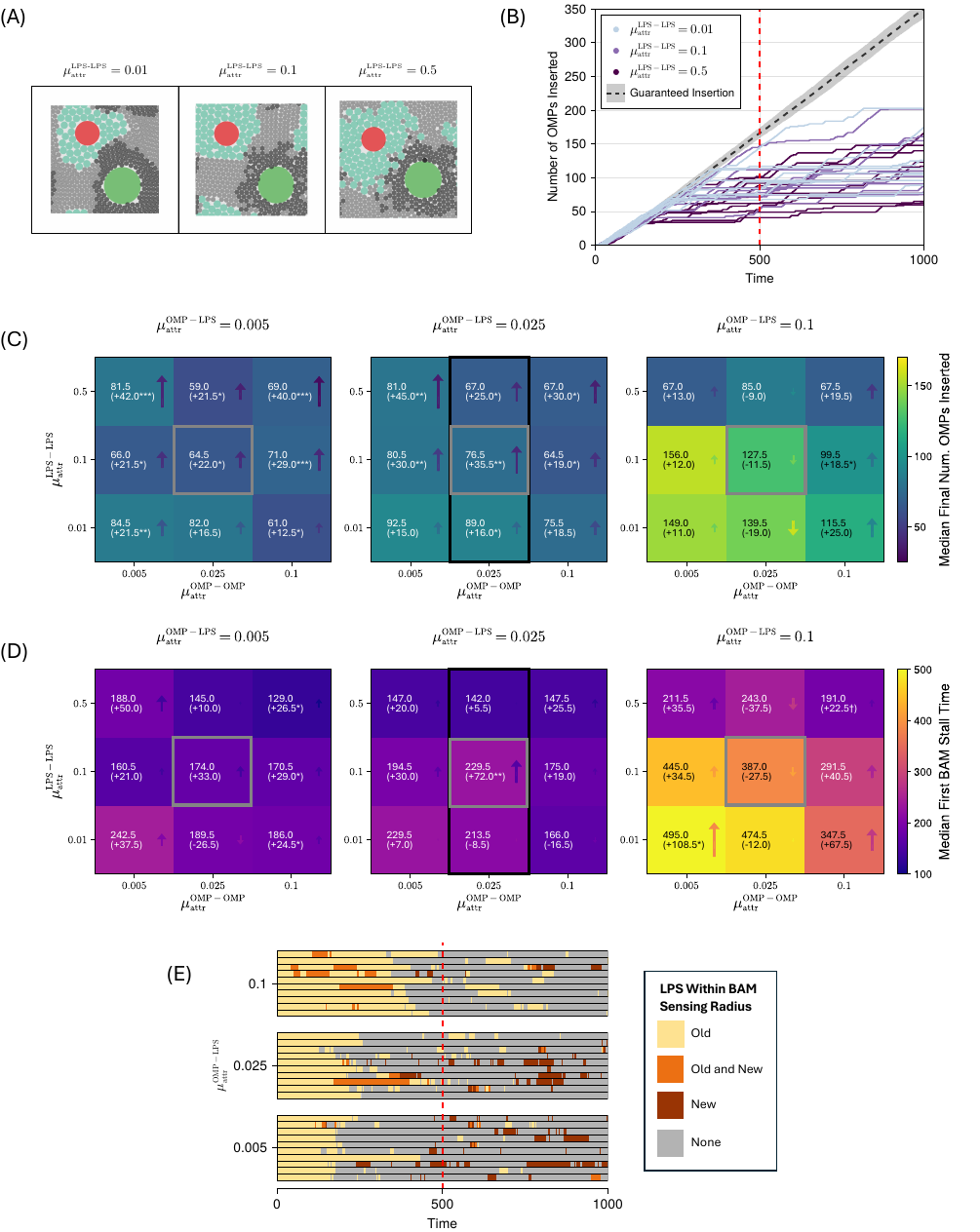}
		\caption{(caption over the page)}
	\end{figure}
	
	\begin{figure}[h!]
		\ContinuedFloat
		\addtocounter{figure}{-1}
		\captionof{figure}{\textbf{Dynamics of a system with BAM and Lpt complexes in abundant LPS.} (A) Representative snapshots of the membrane state as of $t=500$ for a range of values of the LPS-LPS attraction coefficient, $\muattrOMPLPS$. BamA in red, LPS initially present in light grey, LPS inserted during simulation in dark grey, OmpA in pale green, LptD in darker green. (B) Number of inserted OMPs inserted over time (extended to $t=1000$) for the choices of $\muattrLPSLPS$ shown in (A). Trajectories for ten simulations each are shown. We also plot the mean trajectory (dashed black line) and 0.025--0.975 quantiles (shaded grey region) for ten simulations where OMP insertion is not dependent on LPS proximity. (C) Heat map over parameter space of the median final number of OMPs inserted (as at $t=500$) across ten simulations, for combinations of values for the OMP-OMP, OMP-LPS, and LPS-LPS attraction force coefficients ($\muattrOMPOMP$, $\muattrOMPLPS$, $\muattrLPSLPS$, respectively). We also indicate the difference between results from these simulations and results from simulations of a system without an Lpt under the same parameter values (as shown in Supplementary Figure S1). For each parameter combination, we plot an arrow indicating whether the presence of the Lpt increases or decreases the final number of inserted OMPs. Arrow size is proportional to the change in the OMP insertion number; arrow colour corresponds to the colour bar value at the number of OMPs inserted in the BAM-only case. We evaluated the statistical significance of differences in the number of OMPs between the BAM-Lpt and BAM-only systems by two-tailed Mann-Whitney U-test. * indicates $P<0.05$, ** indicates $P<0.005$, *** indicates $P<0.0005$. Black cell outlines indicate parameter combinations shown in (A) and (B); grey cell outlines indicate parameter combinations shown in (E). (D) Heat map over parameter space of the median time at which the BAM complex first became stalled due to a lack of nearby LPS in the BAM-Lpt case, with a comparison to the BAM-only case (simulations were extended up to $t=1000$, where necessary, to determine stall times). Same format as (C). $\dagger$ One simulation replicate in the BAM-only case did not result in a stall within $t=1000$ of simulation time. (E) Type of LPS within the sensing radius of the BAM complex over time for different values of $\muattrOMPLPS$. For each value of $\muattrOMPLPS$, each row of the plot corresponds to a simulation, where the colour indicates whether any old LPS (present at the start of the simulation), new LPS (inserted over the course of the simulation), both, or neither can be found within the sensing radius, $\Rsensins$, of the BAM complex. \label{fig:BAM_vs_Lpt}}
	\end{figure}
	
	Comparing the dynamics of the BAM-Lpt system to the BAM-only system discussed above, we found that the presence of the Lpt complex increased the final number of OMPs inserted, provided OMP-LPS \TWEditMayEighth{affinity} was sufficiently low. In Figure~\ref{fig:BAM_vs_Lpt}(C) we show a heat map of the median number of OMPs inserted over simulations of the BAM-Lpt system (up to $t=500$) across attraction force parameter space, and indicate how this differs from the median number of OMPs inserted in the corresponding BAM-only simulations. For each parameter combination, we used a Mann-Whitney U-test to quantify the statistical significance of any difference in the number of OMPs. Figure~\ref{fig:BAM_vs_Lpt}(C) shows that for the lower two $\muattrOMPLPS$ values tested, the presence of the Lpt substantially increases the number of OMPs inserted during the simulation, but for high $\muattrOMPLPS$, there is no significant difference. Comparing the time of the first BAM stall in the same way --- which we show in Figure~\ref{fig:BAM_vs_Lpt}(D) --- reveals that there is no notable difference in the stall times between the BAM-Lpt and BAM-only systems. This suggests that, for low $\muattrLPSLPS$, the Lpt acts to increase BAM productivity not by delaying the stalling of the BAM complex, but by enabling the BAM complex to be productive \textit{after} having initially stalled. Figure~\ref{fig:BAM_vs_Lpt}(D) also shows that all simulations across all parameter combinations resulted in the BAM complex becoming stalled at least once, with the median time of first stall before $500$ time units in each case.
	
	To confirm that the lack of a substantial increase in BAM productivity for high $\muattrOMPLPS$ was not due to the BAM stall occurring too close to the end of the simulation, we recreated Figure~\ref{fig:BAM_vs_Lpt}(C) after extending all simulations out to $t=1000$. This plot is shown in Supplementary Figure S2, and confirms that, even after doubling the simulation duration, there is still little statistical evidence of a difference in the number of OMPs inserted when $\muattrOMPLPS$ is high.
	
	In order to explain why the presence of an Lpt complex increases BAM productivity when OMP-LPS \TWEditMayEighth{affinity} is low, but has no notable effect when OMP-LPS \TWEditMayEighth{affinity} is high, we examined the LPS within the sensing radius of the BAM in all simulations, and assessed whether or not it was present at the start of the simulation. This is shown over time (extended out to $t=1000$) for a range of values of $\muattrOMPLPS$ in Figure~\ref{fig:BAM_vs_Lpt}(E). Figure~\ref{fig:BAM_vs_Lpt}(E) reveals that BAM rescue events (that is, where an LPS enters the sensing radius of an otherwise stalled BAM complex) are much less frequent when $\muattrOMPLPS$ is high, and are mostly mediated by ``old'' LPS that were present at the start of the simulation. When $\muattrOMPLPS$ is low, BAM rescue events are more common, and are predominantly caused by ``new'' LPS, added over the course of the simulation. These observations suggest two modes of action for the Lpt complex in extending BAM productivity. In low $\muattrOMPLPS$ settings, the Lpt complex directly supplies the BAM complex with new LPS, maintaining its activity over a long period of time (up to $t=1000$) in short bursts, whereas when $\muattrOMPLPS$ is high, new LPS cannot reach the BAM complex, so rescue from the stalled state can only arise from membrane repacking events --- caused by the insertion of new LPS --- which occasionally result in an old LPS ending up near the BAM complex. The former mechanism is likely able to occur when $\muattrOMPLPS$ is low either because the LPS can more easily permeate through the OMP layer to reach the BAM complex, or because there are simply fewer OMPs \TWEditMayTwentyEighth{to traverse in the system} in this case.

	\subsection*{Dynamics of a multiple-BAM system}
	
	We have so far only considered the dynamics of isolated BAM complexes, however, \textit{in vivo}, BAM complexes are known to be relatively highly expressed (around 300 BamA molecules per \textit{E. coli} cell), and gathered in so-called assembly precincts \cite{gunasinghe_et_al_WD40_protein}. To explore the dynamics of a multiple-BAM system, we carried out larger-scale simulations of the model with ten BAM complexes and 2000 LPS initially present. This is equivalent to ten instances of the single-BAM setup outlined in Figures~\ref{fig:LPS-LPS_stickiness_time_series}--\ref{fig:LPS-LPS_clustering} within the same system. We again carried out ten simulations for each of a range of LPS-LPS attraction force coefficients.
	
	Simulating \TWEditMayEighth{this multiple-BAM, assembly precinct} system revealed spatial structure not obvious in single-BAM simulations. Snapshots of the final state of the system, examples of which are shown in Figure~\ref{fig:big_sim_LPS_stickiness}(A), reveal that low LPS-LPS \TWEditMayEighth{affinity} results in a multiple, diffuse OMP clusters, with interstitial LPS trapped among the OMPs, whereas high LPS-LPS \TWEditMayEighth{affinity} leads to the formation of a single OMP cluster which is entirely distinct from the pool of LPS.
	\begin{figure}[h!]
		\centering
		\includegraphics[width=\textwidth, page=6, trim={2cm, 6.2cm, 2cm, 1.9cm}, clip]{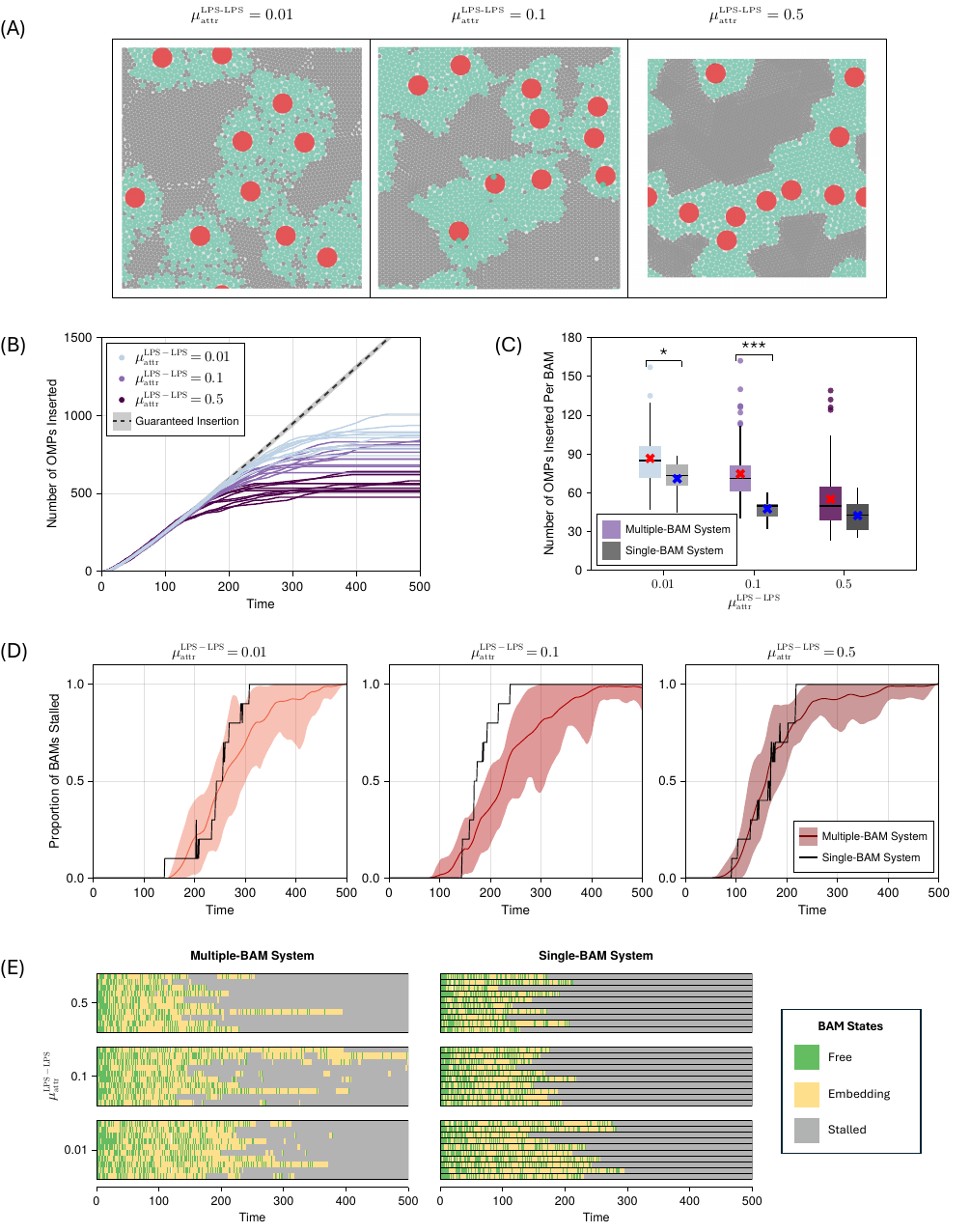}
		\caption{(caption over the page)}
		\end{figure}
		
		\begin{figure}[h!]
		\ContinuedFloat
		\addtocounter{figure}{-1}
		\captionof{figure}{\textbf{Dynamics of a system containing ten BAM complexes and abundant LPS.} (A) Representative snapshots of the membrane state as of $t=500$ for different values of the LPS-LPS attraction force ($\muattrLPSLPS$). As in Figure~\ref{fig:LPS-LPS_stickiness_time_series}: BamA in red, LPS in grey, OmpA in pale green. (B) Number of OMPs inserted over time for different values of the LPS-LPS attraction force, $\muattrLPSLPS$. We show trajectories from ten simulations for each value of $\muattrLPSLPS$. We also plot the mean trajectory (dashed black line) and 0.025--0.975 quantiles (shaded grey region) for ten simulations where OMP insertion is not dependent on LPS proximity. (C) Distribution of the final number of OMPs inserted per BAM complex (as at $t=500$) for each of the choices of $\muattrLPSLPS$ in (A) and (B), compared with the number inserted in a single-BAM system with the same parameters (as in Figure~\ref{fig:LPS-LPS_stickiness_time_series}). Crosses indicate means. We evaluated the statistical significance of differences in the number of OMPs between the multiple- and single-BAM systems by two-tailed Mann-Whitney U-test. * indicates $P<0.05$, ** indicates $P<0.005$, *** indicates $P<0.0005$. (D) Proportion of BAM complexes in the \textit{stalled} state over time (mean and 0.025--0.975 quantiles over ten simulations) for each of the choices of $\muattrLPSLPS$ in (A)--(C). We also show the probability over time of the (single) BAM complex being in the stalled state for the single-BAM system. Curves for multiple-BAM system smoothed using loess smoothing. (E) BAM states over time for both the multiple-BAM and single-BAM systems for each of the values of $\muattrLPSLPS$ in (A)--(D). For each value of $\muattrLPSLPS$, each row corresponds to a single BAM complex; in the multiple-BAM case we show data from all BAM complexes in a single simulation replicate. \label{fig:big_sim_LPS_stickiness}}
	\end{figure}
	
	As with the other simulation setups explored in this work, all ten BAM complexes became stalled (at least at some point) in all simulations of the multiple-BAM system, regardless of the \TWEditMayEighth{choice} of LPS-LPS \TWEditMayEighth{attraction force coefficient}. This can be seen in Figure~\ref{fig:big_sim_LPS_stickiness}(B), where we plot the time series for the number of OMPs inserted across all simulations. While the presence of multiple BAMs has a smoothing effect on the total number of OMPs inserted over time compared to the abrupt cut-off seen in, for example, Figure~\ref{fig:LPS-LPS_stickiness_time_series}(B), all trajectories can be observed to eventually plateau in the multiple-BAM simulations. As we saw in previous results, higher LPS-LPS \TWEditMayEighth{affinity} caused the BAM complexes to stall earlier.
	
	We observed that the productivity of each BAM complex is increased when multiple complexes are present. This can be seen in Figure~\ref{fig:big_sim_LPS_stickiness}(C), where we compare the distributions for the number of OMPs inserted per BAM complex in single- and multiple-BAM simulations. Figure~\ref{fig:big_sim_LPS_stickiness}(C) shows that BAM complexes in multiple-BAM systems on average insert more OMPs over the course of a simulation regardless of the choice of LPS-LPS \TWEditMayEighth{attraction force coefficient}. There are two contributing factors to this increased productivity. Firstly, we found that BAM complexes tended to become stalled later in multiple-BAM simulations compared to the single-BAM simulations. This can be seen in Figure~\ref{fig:big_sim_LPS_stickiness}(D), where we compare the proportion of BAM complexes in the stalled state over time for both the single-BAM and multiple-BAM simulations. Figure~\ref{fig:big_sim_LPS_stickiness}(D) shows that BAM complexes in multiple-BAM systems consistently become stalled later than those in single-BAM systems, particularly for lower LPS-LPS \TWEditMayEighth{affinity}. Secondly, we also found that while stalled BAM complexes in single-BAM systems always remained stalled for the remainder of the simulation (there is a slight caveat to this, which we discuss below), having multiple BAM complexes in the system was sufficient to allow stalled complexes to occasionally re-enter a productive state. This is shown in Figure~\ref{fig:big_sim_LPS_stickiness}(E), where we plot BAM states (\textit{free}, \textit{embedding}, or \textit{stalled}; refer to Methods) over time for both multiple-BAM and single-BAM simulations. Figure~\ref{fig:big_sim_LPS_stickiness}(E) confirms that BAM complexes in multiple-BAM systems can return to a productive state multiple times after stalling, whereas stalling is irreversible in a single-BAM system as long as the membrane packing has relaxed. In Figure~\ref{fig:big_sim_LPS_stickiness}(D), a few very narrow spikes can be seen in the proportion of stalled BAM complexes for the single-BAM simulations, indicating a few instances where BAMs were briefly stalled, then re-entered a productive state. These cases can be understood as BAMs that were close to a stalled state, such that, during a membrane repacking event (following an OMP insertion), any LPS near the BAM complex was briefly pushed out of its sensing radius, then returned as agents rearranged to minimise energy. Since membrane relaxation is fast compared to the growth time scale in the model, these events are rare and brief, and were not detected in earlier analyses. By contrast, in the multiple-BAM system, BAM rescue events are far more frequent, and can be seen even after individual BAM complexes have been stalled for sustained periods of time. This is because any BAM complex in the system that continues to insert OMPs acts as a source of noise for the entire system (similar to the effects noted in the high-diffusion system in Figure~\ref{fig:diffusion_sweep} and the BAM-Lpt system in Figure~\ref{fig:BAM_vs_Lpt}). These OMP insertions force the membrane to locally repack, thereby occasionally restoring access to LPS for nearby stalled complexes. In this sense, BAM complexes that are near each other in the membrane appear to form a positive feedback loop.

	\section*{Discussion}
	
	\TLText{The biogenesis of the outer membrane of Gram-negative bacteria is a \TWEditMayEighth{carefully controlled process} that is essential to the growth of bacterial populations.} Yet despite remarkable advances in both experimental and molecular dynamics approaches, we still lack a detailed understanding of the OM on the time scale of the cell division cycle. To this end, we developed a semi-quantitative agent-based model to explore the dynamics of the OM under the process of OMP and LPS insertion.
	
	One of our primary aims in developing the model was to explore the consequences of a BAM complex requiring nearby LPS in order to insert OMPs, as has been suggested in the literature \cite{doyle_et_al_cryoEM, horne_et_al_lipid_bilayer, peterson_et_al_conserved}. This was represented in the model by considering BamA agents to be \textit{free} prior to engagement with a polypeptide substrate, then, upon encountering a polypeptide, either \textit{embedding} a new nascent OMP into the OM, or becoming \textit{stalled}, depending on whether an LPS is within some sensing radius $\Rsensins$. However, under this assumption, we found that repeated OMP insertions resulted in the formation of a protein array that surrounded the BAM complex, blocking access to LPS and leading to the complex becoming stalled. Indeed, even when modelling a BAM complex in a system containing only LPS, and considering LPS-LPS, OMP-LPS and OMP-OMP affinity parameters over several orders of magnitude, we found that the BAM complex became stalled with near certainty (a single simulation replicate did not stall within the simulated time frame), having inserted at most around 160 OMPs. Even when repeating this simulation setup with the addition of explicit LPS diffusion, an Lpt complex, or multiple BAM complexes, all BAM complexes modelled ended up stalling within a similar time frame, having inserted a similar number of OMPs. Considering that the \textit{E. coli} outer membrane contains around 300 BAM complexes \cite{lithgow_et_al_surveying} and at least 10$^6$ OmpA proteins alone \cite{horne_et_al_lipid_bilayer} (plus many other OMP species which are also assembled by BAM), and that we modelled the somewhat artificial initial condition of BAM surrounded only by LPS, our simulations would appear to show insufficient BAM productivity to account for the quantity of OMPs in the OM.
	
	While none of the parameter schemes or model configurations we explored were sufficient to prevent BAM stalling, we did identify multiple conditions which extended the productive period of the BAM complex. For one, we found that parameters settings which reduced LPS-LPS or OMP-OMP affinity or increased OMP-LPS affinity acted to delay the time at which the BAM complex became stalled in the model. This was because under these conditions the ``stickiness'' between agents of the same type (relative to the between-type stickiness) is reduced, increasing the extent to which their populations can intermix and thus the extent to which LPS can penetrate the OMP array to access the BAM complex. Likewise, the addition of explicit LPS diffusion in the model was also found to extend the productive period of the BAM complex, but for a subtly different reason: the introduction of noise in the system facilitated more random reshuffling of agents, occasionally permitting LPS to access regions closer to the BAM complex. The inclusion of explicit LPS diffusion also introduced an additional mode of action in increasing BAM productivity, in that this allowed for membrane repacking to occur even after the BAM complex stalled. This occasionally resulted in BAM complexes being rescued from a stalled state, entering a short burst of activity before becoming stalled again. Such dynamics were not possible in isolated-BAM systems without an explicit source of noise.
	
	We also found that BAM complexes were more productive in simulations which also included an Lpt complex or other BAM complexes (and a proportional number of LPS), compared to BAM complexes among LPS only. These additional ``source'' complexes increased BAM productivity in a similar manner to explicit LPS diffusion: by repeatedly adding agents to the membrane, the system was forced to rearrange more frequently, increasing the chance of LPS ending up nearer to the BAM complexes. As was the case when explicit LPS diffusion was present, this also provided a means for BAM complexes to temporarily escape the stalled state. 
	
	In a biological context, this might suggest some synergistic relationships between molecular complexes that may aid in the biogenesis of the OM. For example, we saw that Lpt complexes in the neighbourhood of BAM complexes act to supply the BAM complexes with new LPS. \TLText{The scenario is biologically relevant as the core subunit of the Lpt complex, LptD, is itself a $\beta$-barrel, thus at least some BAM complexes will have inserted a neighbouring LptD.} Moreover, we saw that groups of BAM complexes in close proximity --- consistent with observed BAM assembly precincts \textit{in vivo} \cite{gunasinghe_et_al_WD40_protein, lithgow_et_al_surveying} --- are able to insert more OMPs than BAM complexes in isolation. We suggest that this precinct arrangement enables a positive feedback loop between BAM complexes, increasing the overall rate of OMP incorporation into the membrane. The actual arrival of new BAM or Lpt complexes was not modelled in this work, but would shed further light on the emergence and consequences of these collaborative relationships between complexes.

	 Taken together, our model simulations show a BAM complex which self-isolates from nearby LPS and whose OMP-embedding action is highly prone to stalling. This suggests one of a few possible interpretations. One: that the requirement of LPS in the vicinity of the BAM complex for OMP insertion is too strong. This seems unlikely, since the assumption of LPS in the neighbourhood of the BAM complex \TLText{is consistent with biophysics and molecular dynamics simulations of bacterial outer membranes that make use of interstitial LPS \cite{khalid_et_al_atomistic_and_coarse, im_and_khalid_molecular_simulations, webby_et_al_lipids, gutishvili_et_al_seeing_is_believing, iadanza_et_al_distortion, liu_and_gumbart_membrane_thinning} and since structural analyses of OMPs observe interstitial LPS \cite{efremov_structure}. \TWEditMayEighth{Moreover,} in static representations of the BAM complex engaged with substrate OMPs, it is considered that the insertion of an OMP will induce a localised region of increased membrane tension \cite{doyle_et_al_cryoEM}}, which is unlikely to occur in a stiff, lipid-poor environment \cite{horne_et_al_lipid_bilayer}. Two: that BAM complexes indeed spend much of their time in a stalled state, and insert OMPs in short bursts of activity, potentially following membrane repacking events which release LPS into the neighbourhood of the BAM complex. There is some experimental evidence to support this view, with one early study showing the BAM-mediated insertion of fluorescently-labelled LamB taking place in bursts \cite{ursell_et_al_surface_protein}, however, further experimental investigation is required to confirm such a process and explain its origins. Three: that some biomolecular mechanism or extreme parameter configuration not considered in our analysis would provide the BAM complex with a steady supply of LPS and allow it to remain productive throughout the cell division cycle. While possible, we saw no evidence of our model permitting such dynamics without dramatically increasing the rate at which LPS can infiltrate the OMP array surrounding the BAM complex. Based on our simulations, this seems unlikely to occur without an explicit mechanism to draw OMPs away from the BAM, or the introduction of strong, explicit OMP (as well as LPS) diffusion, which is not supported by experimental observation \cite{lee_upton_et_al_slow_diffusion_BAM, benn_et_al_OmpA_order, horne_et_al_lipid_bilayer}. However, our model does contain a number of simplifications from the \textit{in vivo} biology of the OM, which could influence its emergent structure. This point is discussed below.
	 
	 The main limitation of our study is that our model, in its current state, is only semi-quantitative. In order to provide more precise and quantitative predictions of OM dynamics, a detailed parameter estimation study against experimental \TWEditMayTwentyEighth{data} and MD \TWEditMayTwentyEighth{simulations} will be necessary. The fact that certain parameter combinations are more conducive to BAM productivity might suggest that such values are more closely aligned with the biological reality, but this is not necessarily obvious. In particular, we consistently saw that parameter combinations which increased BAM productivity also increased the proportion of LPS molecules which bordered an OMP; in other words, the LPS became increasingly well-mixed among the OMP population. This would appear to contradict the spatial distribution of LPS seen under atomic force microscopy, which showed distinct patches of LPS distributed within a lipid-poor protein array \cite{benn_et_al_OmpA_order, lithgow_et_al_surveying}. Rigorous parameter estimation is therefore needed to formally and quantitatively assess what parameter values are most applicable to the \textit{in vivo} reality.
	 
	 Fortunately, rich sources of both experimental and synthetic data, as well as established parameter estimation pipelines are available. \TLText{For example, fluorescently labelled OmpA has been used as a reporter of OMP folding, so that single-turnover kinetic parameters could be determined for BAM complexes purified from \textit{E. coli} and reconstituted into liposome membranes \cite{bergman_and_sousa_flourescent}. Systems such as this provide precise measurements of a folding constant ($k_{\tinytext{fold}}$) --- for example, 0.78 $\pm$ 0.15 min$^{-1}$ for OmpA in the experimental system described above --- and the affinity between the BAM complex and the OmpA substrate polypeptide (3.1 $\pm$ 1.1 $\mu$M). \TWEditMayEighth{Translating these values into the context of our model is not necessarily trivial. For example,} a limitation of these biochemical assays in that they depend on non-native (i.e. phospholipid) membrane. \TWEditMayEighth{Nonetheless, these experimental data} extend our understanding of the capability and efficiency of the process of outer membrane growth} and provide a potential avenue for model validation.
	 
	 Previous efforts to develop agent-based modelling for the OM have also directly extracted parameters from MD data. Dunton \textit{et al.}, for instance, used umbrella sampling to construct a potential of mean force curve as a function of distance between NanC monomers, which could then be used as a kind of attraction-repulsion force curve \cite{dunton_thesis}. While this approach offers a direct translation between MD-scale modelling and the agent-based context, umbrella sampling carries a substantial computational cost. It is moreover challenging to ensure sufficient sampling of conformational space, especially in slow-moving LPS-containing membranes (Dunton \textit{et al.} considered NanC in a non-native phospholipid bilayer \cite{dunton_thesis}). Alternatively, Chavent \textit{et al.} \cite{chavent_et_al_nanoscale} and Rassam \textit{et al.} \cite{rassam_et_al_supramolecular} calibrated their agent-based models based on the observed motions of membrane molecules in MD simulations. Compared to umbrella sampling, this approach has a weaker connection to the MD context, but comes at a greatly reduced computational cost. Each group of authors devoted considerable effort to analysing the displacement and effective diffusion patterns, which was used both for validation and to calibrate the model time scale \cite{dunton_thesis, chavent_et_al_MD_simulations, rassam_et_al_supramolecular}. This may be a useful metric by which to compare our model to both MD data and experimental observation. 
	 
	 \TLText{For simplicity, \TWEditMayEighth{in this work we} considered OmpA as the only substrate polypeptide (i.e. OMP) embedded by the BAM complexes. This was reasoned on the observation that OmpA is a highly abundant protein in the \textit{E. coli} outer membrane, so in any given moment a relatively large proportion of the BAM complexes would be embedding OmpA polypeptides. However, the outer membranes of \textit{E. coli} and other bacteria \TWEditMayEighth{also contain a wide variety of other OMPs which we have not modelled here. Most notably among these are the abundant} major porins (in \textit{E. coli}, these are OmpC and OmpF). The structures of these major porins are known: they are homo-trimers, wherein three identical barrels come together through protein-protein contacts that exclude interstitial lipids \cite{cowan_et_al_crystal, basle_et_al_crystal, efremov_structure}. \TWEditMayEighth{The \textit{in vivo} trimerisation process, however, is not well understood, and moreover the} overall trimer is non-cylindrical, \TWEditMayEighth{thus it is not immediately clear how this should be accommodated in our modelling framework. For simplicity, we defer the modelling of the major porins to a future study.}}
	 
	 Another potential limitation of our work is that our simulations show the formation of pure OmpA clusters, whereas a recent joint experimental--MD study suggested that direct OMP-OMP contact in the OM was rare, and usually mediated by interstitial lipid \cite{webby_et_al_lipids}. There are two reasons we have chosen not to actively pursue these dynamics in our model. Firstly, if there truly are 2--4 LPS per OMP, as has been suggested \cite{horne_et_al_lipid_bilayer, lessen_et_al_building}, it would seem from an accounting perspective that there are insufficient LPS in the outer membrane to fully encircle all OMPs, while leaving sufficient LPS to form the lipid patches observed under atomic force microscopy \cite{benn_et_al_OmpA_order, lithgow_et_al_surveying}. It is worth noting that in the study that reported the OMP-LPS-OMP structural motif, neither the flourescence assays nor MD simulations conducted included OmpA proteins, which represent a vast OMP population and which we model here \cite{webby_et_al_lipids}. Secondly, we have sought to keep the governing rules of our model as simple and physically-motivated as possible, and under the assumptions of the model we did not observe the OMP-LPS-OMP structural motif in our simulations in any region of parameter space. It is possible that with more detailed physical mechanisms in the model this may emerge, but this is beyond the scope of the present study. In any case, we suspect that adding a requirement for LPS to mediate OMP-OMP contacts would not change the fundamental process observed here. We found that the main driver of BAM stalling was that the migration of LPS through the OMP array was too slow to supply the BAM complex. We anticipate that insisting on LPS also appearing in the neighbourhood of all inserted OMPs would \textit{increase} rather than decrease the demands on LPS supply near the BAM complex, therefore not preventing the complex from stalling.

	In this work, we introduced an agent-based model to model the dynamics of the Gram-negative outer membrane on the time scale of the cell division cycle. This model offers insights into the production and maintenance of the OM at a temporal and spatial resolution which cannot be accessed by experimental or conventional molecular dynamics approaches. We hope that future development of the model will provide researchers with a powerful tool to assess and generate hypotheses on OM biogenesis.

	\section*{Acknowledgements}
	The authors wish to acknowledge Syma Khalid of the University of Oxford for useful discussions on molecular dynamics modelling, and Cara Press and Chris Stubenrauch of Monash University for their feedback on the model. This work was also supported through numerous discussions with members of the Australian Research Council (ARC) Centre of Excellence in Mathematical Analysis of Cellular Systems (MACSYS; grant number CE230100001).
	
	TW and KJG's research is supported by the MACSYS ARC Centre of Excellence. JMO acknowledges support by the ARC (DP230100380, FT230100352, and DP260100767); TL acknowledges support by the National Health and Medical Research Council (NHMRC) through an NHMRC Investigator Award (2016330); JF acknowledges support by the ARC (FT210100034, CE230100001) and the NHMRC (APP2019093).

	\clearpage
	\bibliographystyle{apalike}

\end{document}


\maketitle
	
	\clearpage
	\section{Supplementary figures}

	\begin{figure}[h!]
		\centering
		\includegraphics[width=\textwidth, page=1, trim={2cm, 11.3cm, 2cm, 2.5cm}, clip]{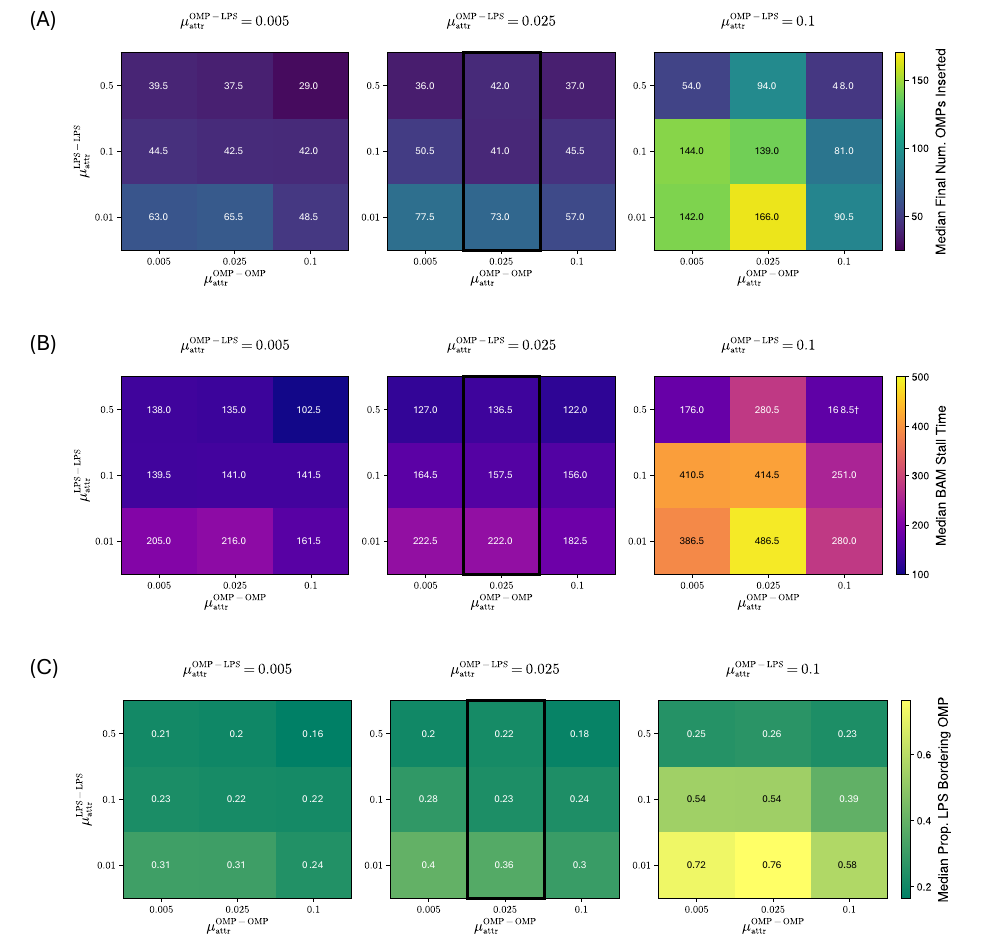}
		\caption{\textbf{Dynamics of an isolated BAM complex in abundant LPS: effect of changing all agent-agent attraction force coefficients.} (A) Heat map of median final number of OMPs inserted (as of $t=500$) across ten simulations for combinations of values for the OMP-OMP, OMP-LPS, and LPS-LPS attraction force coefficients ($\muattrOMPOMP$, $\muattrOMPLPS$, $\muattrLPSLPS$, respectively). The cells with a solid black outline ($\muattrOMPLPS=0.025$, $\muattrOMPOMP=0.025$) are the parameter combinations used in Figure 2 in the main text. (B) Heat map of the median time of the BAM complex stalling (simulations were extended up to $t=1000$, where necessary, to determine stall times); same parameter combinations and format as (A). $\dagger$ One simulation replicate did not result in a stall within $t=1000$ of simulation time. (C) Heat map of the median proportion of LPS bordering an OMP at the end of simulations ($t=500$); same parameter combinations and format as (A) and (B).}
		\label{fig:stickiness_3D}
	\end{figure}
	
	\clearpage

	\begin{figure}[h!]
		\includegraphics[width=\textwidth, page=2, trim={2cm, 22.3cm, 2cm, 2.6cm}, clip]{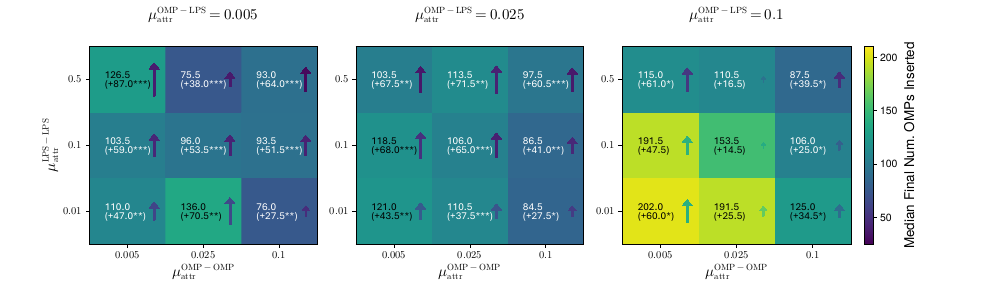}
		\caption{\textbf{Dynamics of a system with BAM and Lpt complexes in abundant LPS (extended).} Heat map of the median final number of OMPs inserted across ten simulations run to $t=1000$, for combinations of values for the OMP-OMP, OMP-LPS, and LPS-LPS attraction force coefficients ($\muattrOMPOMP$, $\muattrOMPLPS$, $\muattrLPSLPS$, respectively). We also indicate the difference between results from these simulations and results from simulations of a system without an Lpt under the same parameter values (as shown in Supplementary Figure~\ref{fig:stickiness_3D}). For each parameter combination, we plot an arrow indicating whether the presence of the Lpt increases or decreases the final number of inserted OMPs. Arrow size is proportional to the change in the OMP insertion number; arrow colour corresponds to the colour bar value at the number of OMPs inserted in the BAM-only case. We evaluated the statistical significance of differences in the number of OMPs between the BAM-Lpt and BAM-only systems by two-tailed Mann-Whitney U-test. * indicates $P<0.05$, ** indicates $P<0.005$, *** indicates $P<0.0005$. \textbf{Note that the colour bar scale for this figure differs to those in the main text and Supplementary Figure~\ref{fig:stickiness_3D}.}}
		\label{fig:BAM_vs_Lpt_1000}
	\end{figure}

	\clearpage

	\section{Model description}
	\label{sec:model_description}
	
	The agent-based model introduced in this work considers a small patch of membrane occupied by two main groups of agents: outer membrane proteins (OMPs) and lipopolysaccharides (LPS). We consider three types of OMP --- OmpA, LptD, and BamA --- and a single type of LPS. We also include in the model populations of nascent OMP and LPS agents --- with their own slightly modified properties --- as they emerge into the membrane. The overall logic of the model is laid out in Supplementary Algorithm~\ref{alg:ABM_algo} using the shorthand notation $\mathcal{P}$ and $\mathcal{L}$ for the populations of OMPs and LPS respectively, $\mathcal{P}_{\tinytext{nasc}}$ and $\mathcal{L}_{\tinytext{nasc}}$ for their corresponding nascent populations, $n_{\tinytext{poly}}$ for the number of polypeptides in the vicinity of the modelled section of membrane (we do not model their position in space), $\Omega$ for the size of the spatial domain modelled, $\Phi$ for the initial conditions, and $\Theta$ for the model parameters.

		\begin{algorithm}
		\caption{\textbf{Overall logic of the membrane ABM}}
		\label{alg:ABM_algo}
		\begin{algorithmic}
			\State \codecomm{randomise initial agent positions, define initial polypeptide number}
			\State  $ \mathcal{P}, \mathcal{L}, n_{\text{\tiny poly}}, \Omega \gets \texttt{initialise_system}(\Phi)$
			\State $\Pnascent \gets \emptyset,~ \Lnascent \gets \emptyset$
			\State
			\State \codecomm{run force resolution on the initial system until stabilised}
			\State $ \mathcal{P}, \mathcal{L} \gets \texttt{equilibrate_system}(\mathcal{P}, \mathcal{L}, \Omega \vert \Theta)$
			\State 
		    \State \codecomm{if (large) holes have formed, iteratively shrink and re-equilibrate}
		    \State $ \mathcal{P}, \mathcal{L}, \Omega \gets \texttt{shrink_domain}(\mathcal{P}, \mathcal{L}, \Omega \vert \Theta)$
			\State 
			\State \codecomm{run the main loop}
			\State $t\gets 0$
			\While{$t<t_{\text{\tiny max}}$}
			\State \codecomm{update BamA states, new OMP insertions, polypeptide arrivals}
			\State $ \mathcal{P}, \Pnascent, n_{\text{\tiny poly}}, \gets \texttt{BAM_subsystem}(\mathcal{P}, \mathcal{L}, \Pnascent, n_{\text{\tiny poly}}, \Omega, t \vert \Theta)$
			\State
			\State \codecomm{update LptD states, new LPS insertions}
			\State $ \mathcal{P}, \Lnascent \gets \texttt{Lpt_subsystem}(\mathcal{P}, \Lnascent, \Omega, t \vert \Theta)$
			\State
			\State \codecomm{update nascent agents}
			\State $ \mathcal{P}, \mathcal{L}, \Pnascent, \Lnascent \gets \texttt{update_nascent_agents}(\mathcal{P}, \mathcal{L}, \Pnascent, \Lnascent, \Omega, t \vert \Theta)$
			\State
			\State \codecomm{rescale domain and agent positions to account for insertion of new agents}
			\State $ \mathcal{P}, \mathcal{L}, \Pnascent, \Lnascent, \Omega \gets \texttt{rescale_domain}(\mathcal{P}, \mathcal{L}, \Pnascent, \Lnascent, \Omega \vert \Theta)$
			\State 
			\State \codecomm{simulate diffusion of agents}
			\State $ \mathcal{P}, \mathcal{L}, \Pnascent, \Lnascent \gets \texttt{agent_diffusion}(\mathcal{P}, \mathcal{L}, \Pnascent, \Lnascent, \Omega \vert \Theta)$
			\State
			\State \codecomm{resolve forces}
			\State $ \mathcal{P}, \mathcal{L}, \Pnascent, \Lnascent \gets \texttt{resolve_forces}(\mathcal{P}, \mathcal{L}, \Pnascent, \Lnascent, \Omega \vert \Theta)$
			\State
			\State $t\gets t+ \Delta t$
			\EndWhile
		\end{algorithmic}
		\end{algorithm}
		
		In the following subsection we introduce the components of the model, as well as the formal notation to be used in the full model definition, then in later subsections give full details of the submodels listed in Supplementary Algorithm~\ref{alg:ABM_algo}. The general logic of the initialisation process is shown in Supplementary Figure~\ref{fig:initialisation}.

		\begin{figure}[h!]
			\centering
			\includegraphics[width=\textwidth, page=3, trim={2cm, 23cm, 2cm, 2.6cm}, clip]{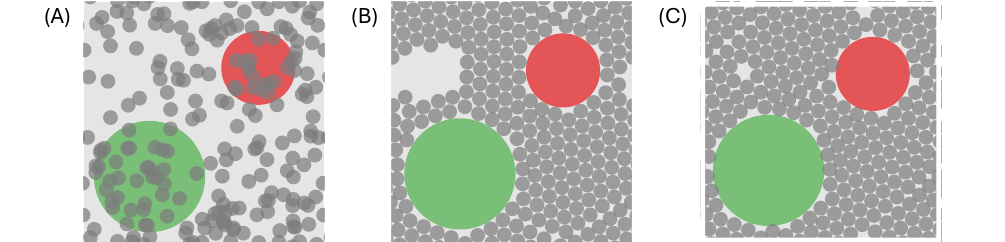}
			\caption{Initialisation of the system. (A) We first place all agents randomly within the domain (described in Supplementary Section~\ref{sec:initialisation}). (B) Next, we equilibrate the system by running force resolution over some equilibration time (described in Supplementary Section~\ref{sec:equilibration}). (C) If holes are present in the domain, we iteratively shrink the domain and re-equilibrate until no sufficiently large holes remain (described in Supplementary Section~\ref{sec:shrink_wrapping}).}
			\label{fig:initialisation}
		\end{figure}

		\subsection{Model components and notation}
		
		Here, we describe the components of the model and introduce the notation which will be used in the remainder of the section.
		
		\begin{itemize}
		\item \textbf{Domain:} $\Omega(t)\in\mathbb{R}^2$ is the spatial domain of the section of membrane modelled, and grows dynamically with the number of agents. We define $\Omega(t)$ as the rectangular region $[0,X(t)]\times[0,Y(t)]$, and apply periodic boundary conditions in the $x$- and $y$-directions.
		\item \textbf{Agents:} Each ``full'' agent has a unique index $i\in I_{\tinytext{full}}(t)\subset \mathbb{N}$ and a type \\$\gamma(i)\in \Gamma=\{\text{LPS},\text{OmpA},\text{LptD},\text{BamA}\}$, and is described by some circle with centre $\mathbf{r}^{(i)}(t)\in\Omega(t)$ and radius $R_{\gamma(i)}$. We denote the set of indices belonging to (full) agents of type $\gamma^*\in\Gamma$ at time $t$ as $I_{\gamma^*}(t) = \{i~:\gamma(i)=\gamma^*~\forall i \in I_{\tinytext{full}}\}$. We denote the set of \textbf{all} agent indices (including nascent agents, introduced below) as $I_{\tinytext{all}}(t)=I_{\tinytext{full}}(t)\cup I_{\tinytext{nasc}}(t)$.
		\item \textbf{BamA and LptD:} BamA and LptD agents are each assumed to be representative of assembled BAM and Lpt complexes, respectively, which can insert new OMP and LPS agents, respectively, into the system during model simulations. To track this, these agents have an ``insertion state'' which changes dynamically as the system evolves. The insertion state of each BamA $i$ at time $t$ is given by $S^{(i)}(t)\in\{\textit{free}, \textit{embedding}, \textit{stalled}\}$, where the \textit{free} state indicates that the BamA is waiting to bind with a polypeptide, the \textit{embedding} state indicates that the BamA is inserting a new OMP into the membrane, and the \textit{stalled} state indicates that the BamA is bound to a polypeptide but is unable to insert it, as there are no LPS within the sensing radius $R^{\tinytext{sens}}$ of the BamA (full details on the OMP insertion model in Supplementary Section~\ref{sec:nascent_agent_insertion}). The insertion state of LptD $i$ at time $t$ is given by $S^{(i)}(t)\in\{\textit{free}, \textit{embedding}\}$. Note that LptDs do not enter a \textit{stalled} state since the Lpt is assumed not to depend on nearby LPS (full details of the LPS insertion model in Supplementary Section~\ref{sec:nascent_agent_insertion}).
		\item \textbf{Nascent agents:} OMPs and LPS also have nascent variants, which are the sets of agents in the process of being inserted, but which are not yet fully embedded in the membrane (full details of the insertion models in Supplementary Section~\ref{sec:nascent_agent_insertion}). In this work, we only consider nascent OmpA and LPS. Like regular agents, nascent OMPs and LPS each have a unique index $i \in I_{\tinytext{nasc}}(t) \subset \mathbb{N}$ (such that $I_{\tinytext{nasc}}(t)\cap I_{\tinytext{full}}(t) = \emptyset$), a type $\gamma(i)\in\Gamma$, and are also described by circles characterised by a position $\mathbf{r}^{(i)}(t)\in \Omega(t)$ and radius $R_{\gamma(i)}$. We also track each nascent agent's ``arrival time'', that is, the time at which each nascent agent first begun to be inserted, and the index of the agent inserting it. The arrival time of nascent agent $i$ is given by $t_{\tinytext{arr}}^{(i)}$, and the index of its inserting agent is $a_{\tinytext{ins}}^{(i)}$.
		\item \textbf{Substrate polypeptides:} $n_{\text{{\tiny poly}}}$ is the number of substrate polypeptides in the vicinity of the model membrane, used to determine when BamA agents leave the \textit{free} state and enter the \textit{embedding} or \textit{stalled} states. We do not model the explicit positions of substrate polypeptides.
		\end{itemize}

		\subsection{Submodels}
		
		\subsubsection{Initialisation}
		\label{sec:initialisation}
		In this component of the model, we account for initialisation of the system, which we do by simply assigning random positions in the domain to each agent. This represents \texttt{initialise_system} in Supplementary Algorithm~\ref{alg:ABM_algo}.
		
		Assume for ease of notation that the initial population of agents has a set of unique indices $I_{\tinytext{full}}^{\tinytext{init}}$ (for example, $\{1, 2, ..., n_{\tinytext{agents}}^{\tinytext{init}}\}$ where $n_{\tinytext{agents}}^{\tinytext{init}}$ is the initial number of agents), and that there are no initially nascent agents.
		
		The initialisation procedure is described in Supplementary Algorithm~\ref{alg:initialise}.
		
		\begin{algorithm}
			\caption{\texttt{initialise_system}}
			\label{alg:initialise}
			\begin{algorithmic}
				\State \codecomm{set initial positions}
				\For{$i\in I_{\tinytext{full}}^{\tinytext{init}}$}
				\State Draw $x\sim\text{unif}(0,X)$, $y\sim\text{unif}(0, Y)$
				\State $\mathbf{r}^{(i)}\gets(x,y)$
				\EndFor
			\end{algorithmic}
		\end{algorithm}

		\subsubsection{Equilibration}
		\label{sec:equilibration}
		In this component of the model, we account for the initial round of equilibration applied to the agents in order to find a stable packing. 
		
		After placing all agents randomly within the domain, we allow the agents to reorganise to find a local energy minimum. We do so by simulating the motion of the agents under the action of the inter-agent attraction-repulsion forces (inter-agent forces are defined in Supplementary Section~\ref{sec:forces}). We simulate force resolution on the time interval $[0,t_{\tinytext{eqm}}]$ for some equilibration time $t_{\tinytext{eqm}}$ with a discrete time step of $\Delta t$. To avoid overcrowding of the domain, we ensure a generous initial domain size; excess area is eliminated in the next stage of initialisation (Supplementary Section~\ref{sec:shrink_wrapping}).
		
		The initial equilibration procedure is described in Supplementary Algorithm~\ref{alg:equilibration}.
		
		\begin{algorithm}
			\caption{\texttt{equilibrate_system}}
			\label{alg:equilibration}
			\begin{algorithmic}
				\State \codecomm{step time over equilibration period}
				\State $\tau \gets 0$
				\While{$\tau< t_{\text{\tiny eqm}}$}
				\State \codecomm{sum attraction-repulsion forces for each agent}
				\For{$i \in I_{\tinytext{all}}$}
				\State $\mathbf{F}^{(i)} \gets \mathbf{0}$
				\For{$j \in I_{\tinytext{all}} \setminus \{i\}$}
				\State Compute $\mathbf{F}_{\tinytext{attr-rep}}^{(ij)}$ (given by Supplementary Equation~\eqref{eq:forces})
				\State $\mathbf{F}^{(i)} \gets \mathbf{F}^{(i)} + \mathbf{F}_{\tinytext{attr-rep}}^{(ij)}$
				\EndFor
				\State \codecomm{update position}
				\State $\mathbf{r}^{(i)} \gets \mathbf{r}^{(i)}(t) + \Delta t/(\eta R_{\gamma(i)})\mathbf{F}^{(i)}$
				\State \codecomm{apply periodic boundaries}
				\State $\mathbf{r}^{(i)} \gets (\mathbf{r}^{(i)}\cdot\hat{\mathbf{e}}_x  \mod X, \mathbf{r}^{(i)}\cdot\hat{\mathbf{e}}_y  \mod Y)$
				\EndFor
				\State $\tau\gets \tau + \Delta t$
				\EndWhile
			\end{algorithmic}
		\end{algorithm}

		\subsubsection{Domain shrinkage}
		\label{sec:shrink_wrapping}
		In this component of the model, we account for any large unoccupied regions of the domain --- that is, holes --- that may have appeared in the domain during equilibration in instances where the emergent agent packing is more dense than is needed to fill the domain. We do so by iteratively shrinking the domain and re-equilibrating until no sufficiently large holes remain.
		
		For each shrinkage step, we first check if a sufficiently large hole can be found in the domain. As in Supplementary Section~\ref{sec:domain_growth}, we define such a hole as a circle of radius $R^{\tinytext{hole}}$ within the domain that has no intersection with any of the agents. If such a hole can be found, we proceed with domain shrinkage. We do so by resizing the domain and dilating the position of all agents by a fixed scaling factor $\sigma_{\tinytext{shrink}}\in (0,1)$. Following this, we then apply a round of equilibration to allow the agents to return to a locally energy-minimal state, as described in Supplementary Section~\ref{sec:equilibration} (except over a shorter time, $t_{\tinytext{eqm short}}$, since the agent positions are already close to equilibrium). We iterate through successive steps until holes considered to be of significant size can no longer be found in the domain.
		
		The domain shrinkage procedure is described in Supplementary Algorithm~\ref{alg:domain_shrinkage}.

		\begin{algorithm}
			\caption{\texttt{shrink_domain}}
			\label{alg:domain_shrinkage}
			\begin{algorithmic}
				\State \codecomm{check if there is a hole}
				\While{$\exists \mathbf{x}^* \in \Omega~\text{s.t.} ~ d_\Omega \left( \mathbf{x}^*, \mathbf{r}^{(i)} \right) > R_{\gamma(i)} + R^{\tinytext{hole}} ~\forall i \in I_{\tinytext{all}}$}
				\State \codecomm{rescale domain and all agent positions}
				\For{$i \in I_{\tinytext{all}}$}
				\State $\mathbf{r}^{(i)} \gets \sigma_{\tinytext{shrink}} \mathbf{r}^{(i)}$
				\EndFor
				\State $\Omega \gets \sigma_{\tinytext{shrink}} \Omega$
				\State
				\State \codecomm{re-equilibrate the system}
				\State $\tau \gets 0$
				\While{$\tau< t_{\text{\tiny short eqm}}$}
				\State \codecomm{sum attraction-repulsion forces for each agent}
				\For{$i \in I_{\tinytext{all}}$}
				\State $\mathbf{F}^{(i)} \gets \mathbf{0}$
				\For{$j \in I_{\tinytext{all}} \setminus \{i\}$}
				\State Compute $\mathbf{F}_{\tinytext{attr-rep}}^{(ij)}$ (given by Supplementary Equation~\eqref{eq:forces})
				\State $\mathbf{F}^{(i)} \gets \mathbf{F}^{(i)} + \mathbf{F}_{\tinytext{attr-rep}}^{(ij)}$
				\EndFor
				\State \codecomm{update position}
				\State $\mathbf{r}^{(i)} \gets \mathbf{r}^{(i)}(t) + \Delta t/(\eta R_{\gamma(i)})\mathbf{F}^{(i)}$
				\State \codecomm{apply periodic boundaries}
				\State $\mathbf{r}^{(i)} \gets (\mathbf{r}^{(i)}\cdot\hat{\mathbf{e}}_x  \mod X, \mathbf{r}^{(i)}\cdot\hat{\mathbf{e}}_y  \mod Y)$
				\EndFor
				\State $\tau\gets \tau + \Delta t$
				\EndWhile
				\EndWhile
			\end{algorithmic}
		\end{algorithm}

		\subsubsection{BAM subsystem}
		\label{sec:BAM_subsystem}
		
		In this component of the model, we account for the action of the BAM complexes, that is, we model changes in the insertion states of the BamAs, create new nascent OMPs as necessary, and update the polypeptide count. This represents $\texttt{BAM_subsystem}$ in Supplementary Algorithm~\ref{alg:ABM_algo}.
		
		Each BamA $i$ in the \textit{free} state at time $t$ can become bound to a polypeptide within the time step $\Delta t$ with probability
		\begin{equation}
			P(S^{(i)}(t+\Delta t)\neq\textit{free}~ \vert S^{(i)}(t)=\textit{free}) = 1 - \exp \left(-\beta \frac{n_{\text{\tiny poly}}}{|\Omega(t)|}\Delta t \right),\quad \forall i\in I_{\tinytext{BamA}}(t),
			\label{eq:free_to_bound}
		\end{equation}
		
		\noindent where $\beta$ is the binding rate of polypeptide to BAM complex. If a binding event occurs, $n_{\text{\tiny poly}}$ is also decremented by 1.
		
		Once bound to a polypeptide, BamA $i$ enters the \textit{embedding} state if there is an LPS agent within some sensing radius $R^{\text{{\tiny sens}}}$ of the BamA. Specifically, there must exist some $j\neq i$ such that $d_{\Omega(t)}(\mathbf{r}^{(i)}(t), \mathbf{r}^{(j)}(t) ) < R_{\text{{\tiny BamA}}} + R_{\text{{\tiny LPS}}} + R^{\text{{\tiny sens}}}$, where $d_{\Omega(t)}(\cdot,\cdot)$ is the Euclidean distance between two points in $\Omega(t)$, using the minimum image convention to account for the periodic boundary conditions. If no there is no LPS within the sensing radius of the BamA at the time it binds to a polypeptide, the BamA instead enters the \textit{stalled} state, where it remains until there is a sufficiently nearby LPS, at which point it switches to the \textit{embedding} state.
		
		Once a BamA $i$ enters the \textit{embedding} state, it creates a new nascent OMP, which is gradually added to the system over some insertion time (the insertion process is described in Supplementary Section~\ref{sec:nascent_agent_insertion}). In this work, we only consider insertion of OmpAs, although the following can easily be extended to accommodate the insertion of other OMP types. New nascent OmpAs are initially added at a random point on the boundary of the BamA. That is, we first select a new index $i^*\in \mathbb{N} \setminus I_{\tinytext{all}}(t)$, draw a random $\theta_{\tinytext{OMP}} \sim \text{unif}(0,2\pi)$, and set the initial position of the new nascent OmpA as 
		\begin{align}
			\mathbf{r}^{(i^*)}(t) = \bigg(&\mathbf{r}^{(i)}(t)\cdot \hat{\mathbf{e}}_x + (R_{\text{{\tiny BamA}}} - R_{\text{{\tiny OmpA}}})\cdot\cos(\theta_{\tinytext{OMP}}) \qquad\text{mod } X(t), \nonumber\\
			&\mathbf{r}^{(i)}(t)\cdot \hat{\mathbf{e}}_y + (R_{\text{{\tiny BamA}}} - R_{\text{{\tiny OmpA}}})\cdot\sin(\theta_{\tinytext{OMP}}) \qquad\text{mod } Y(t)\bigg).
			\label{eq:nascent_OMP_position}
		\end{align}
		
		\noindent Note that the radius of OmpA is less than that of BamA, so at the time of first insertion, the nascent OmpA is entirely enclosed within the inserting BamA.
		
		BamAs remain in the \textit{embedding} state for the fixed period of time it takes to fully insert the nascent OMP. Once the insertion time has elapsed, the BamA is returned to the \textit{free} state. This process is managed by the nascent agent update step described in Supplementary Section~\ref{sec:nascent_agent_insertion}.
		
		Finally, the arrival of new polypeptides is modelled as a Poisson process. If the polypeptide arrival rate is $s_{\tinytext{poly}}$, then the number of new polypeptides added in a given time step is given by
		\begin{equation}
			n_{\text{\tiny poly}}^{\text{\tiny new}} \sim \text{Pois}(s_{\tinytext{poly}} \Delta t).
		\end{equation}
		
		The overall BAM subsystem is described in Supplementary Algorithm~\ref{alg:BAM_subsys}. We also illustrate the action of the BAM subsystem in Supplementary Figure~\ref{fig:BAM_subsystem}.
		
		\begin{figure}[h!]
			\centering
			\includegraphics[width=\textwidth, page=4, trim={2cm, 24cm, 2cm, 2.5cm}, clip]{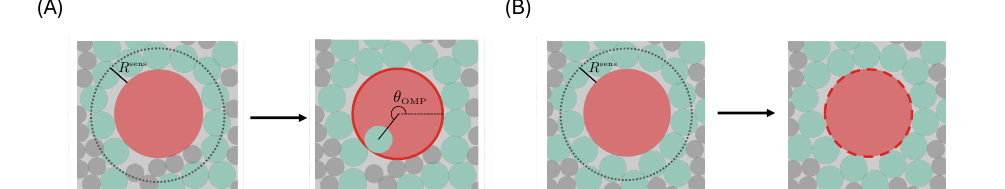}
			\caption{BAM subsystem. Once a BAM complex in the \textit{free} state becomes bound to a polypeptide (with probability governed by Supplementary Equation~\eqref{eq:free_to_bound}), one of the following will take place. (A) If there is an LPS (grey) within the sensing radius $R^{\tinytext{sens}}$ (dotted line) of the free BAM complex (red, no outline), it enters the \textit{embedding} state (solid outline). At this point, a nascent OMP (pale green) is created at a random point on the circumference of the BAM complex (forming angle $\theta_{\tinytext{OMP}}$ to the positive $x$-direction). (B) If there are no LPS within $R^{\tinytext{sens}}$ of the BAM complex, the complex enters the \textit{stalled} state (dashed outline) until an LPS comes within the sensing radius of the BAM complex.}
			\label{fig:BAM_subsystem}
		\end{figure}

			\begin{algorithm}
			\caption{\texttt{BAM_subsystem}}
			\label{alg:BAM_subsys}
				\begin{algorithmic}
					\State \codecomm{iterate over BamAs}
					\For{$i \in I_{\tinytext{BamA}}$}
					\State \codecomm{check for polypeptides binding to free BamAs}
					\If{$S^{(i)}=\textit{free}$}
					\State Draw $x\sim \text{unif}(0,1)$
					\If{$x<1-\exp(-\beta (n_{\text{\tiny poly}}/|\Omega|)\Delta t)$}
					\State $S^{(i)} \gets \textit{stalled}$ \codecomm{temporarily assign stalled state}
					\State $n_{\text{\tiny poly}} \gets n_{\text{\tiny poly}} - 1$
					\EndIf
					\EndIf
					\State \codecomm{check all stalled BamAs for nearby LPS}
					\State \codecomm{(includes BamAs which bound to a polypeptide this time step)}
					\If{$S^{(i)}=\textit{stalled}$}
					\If{$\exists j \in I_{\tinytext{LPS}}$ s.t. $d_{\Omega}(\mathbf{r}^{(i)}, \mathbf{r}^{(j)}) < R_{\text{{\tiny BamA}}} + R_{\text{{\tiny LPS}}} + R^{\text{\tiny sens}}$}
					\State $S^{(i)}\gets \textit{embedding}$
					\State \codecomm{create new nascent OmpA}
					\State Select $i^* \in \mathbb{N} \setminus I_{\tinytext{all}}$
					\State $I_{\tinytext{nasc}} \gets I_{\tinytext{nasc}}\cup \{i^*\}$
					\State Draw $\theta_{\tinytext{OMP}} \sim \text{unif}(0, 2\pi)$
					\State Set $\mathbf{r}^{(i^*)}$ using Supplementary Equation~\eqref{eq:nascent_OMP_position}
					\State \codecomm{record arrival time, inserting agent index}
					\State $t_{\tinytext{arr}}^{(i^*)} \gets t$
					\State $a_{\tinytext{ins}}^{(i^*)}\gets i$
					\EndIf
					\EndIf
					\EndFor
					\State
					\State \codecomm{check for new polypeptide arrivals}
					\State Draw $n_{\text{\tiny poly}}^{\text{\tiny new}} \sim \text{Pois}(s_{\tinytext{poly}} \Delta t)$
					\State $n_{\text{\tiny poly}} \gets n_{\text{\tiny poly}} + n_{\text{\tiny poly}}^{\text{\tiny new}}$
				\end{algorithmic}
		\end{algorithm}

		\subsubsection{Lpt subsystem}
		\label{sec:Lpt_subsystem}
		
		In this component of the model, we account for the actions of the Lipopolysaccharide transport (Lpt) complexes. The Lpt subsystem --- responsible for LPS insertion in the outer membrane --- is modelled as a similar process to the BAM subsystem, with the simplification that we do not assume LPS insertion is dependent on local membrane composition. This component of the model represents \texttt{Lpt_subsystem} in Supplementary Algorithm \ref{alg:ABM_algo}.
		
		Each LptD $i$ in the \textit{free} state at time $t$ can begin \textit{embedding} a nascent LPS within the time step $\Delta t$ with probability
		\begin{equation}\label{eq:Lpt_free_to_embedding}
			P(S^{(i)}(t+\Delta t)=\textit{embedding} ~\vert S^{(i)}(t)=\textit{free}) = 1 - \exp(-s_{\tinytext{LPS}}\Delta t),\quad \forall i\in I_{\tinytext{LPS}}(t)
		\end{equation}
		
		\noindent where $s_{\tinytext{LPS}}$ is the arrival rate of new LPS (since $s_{\tinytext{LPS}}\Delta t$ is very small in practice, this is approximately equivalent to assuming Poisson arrivals of new nascent LPS).
		
		As with the BamA subsystem, once an LptD enters the \textit{embedding} state, it creates a new nascent LPS, which is gradually added to the system over some insertion time (the insertion process is described in Supplementary Section~\ref{sec:nascent_agent_insertion}). New nascent LPS are initially added at a random point on the boundary of the LptD. That is, we first select a new index $i^*\in \mathbb{N} \setminus I_{\tinytext{all}}(t)$, draw a random $\theta_{\tinytext{LPS}} \sim \text{unif}(0,2\pi)$, and set the initial position of the new nascent LPS as 
		\begin{align}
			\mathbf{r}^{(i^*)}(t) = \bigg(&\mathbf{r}^{(i)}(t)\cdot \hat{\mathbf{e}}_x + (R_{\text{{\tiny LptD}}} - R_{\text{{\tiny LPS}}})\cdot\cos(\theta_{\tinytext{LPS}}) \qquad\text{mod } X(t), \nonumber\\
			&\mathbf{r}^{(i)}(t)\cdot \hat{\mathbf{e}}_y + (R_{\text{{\tiny LptD}}} - R_{\text{{\tiny LPS}}})\cdot\sin(\theta_{\tinytext{LPS}}) \qquad\text{mod } Y(t)\bigg).
			\label{eq:nascent_LPS_position}
		\end{align}
		
		\noindent Like with the BamA subsystem, the radius of LPS is much smaller than that of LptD, so at the time of first insertion, the nascent LPS is entirely enclosed within the inserting LptD.
		
		LptDs remain in the \textit{embedding} state for the fixed period of time it takes to fully insert the nascent LPS, after which it returns to the \textit{free} state. This process is managed by the nascent agent update step described in Supplementary Section~\ref{sec:nascent_agent_insertion}.

		The overall Lpt subsystem is described in Supplementary Algorithm~\ref{alg:Lpt_subsys}. We also illustrate the action of the Lpt subsystem in Supplementary Figure~\ref{fig:Lpt_subsystem}.

		\begin{figure}[h!]
			\centering
			\includegraphics[width=\textwidth, page=5, trim={2cm, 24.5cm, 2cm, 2.6cm}, clip]{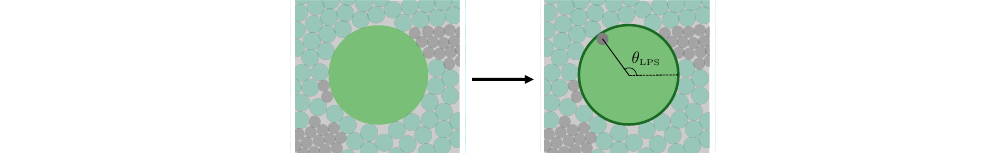}
			\caption{Lpt subsystem. An LptD in the \textit{free} state can enter the \textit{embedding} state (solid outline) according to the transition probability in Supplementary Equation~\eqref{eq:Lpt_free_to_embedding}. At this point, a nascent LPS is created at a random point on the circumference of the LptD (forming angle $\theta_{\tinytext{LPS}}$ to the positive $x$-direction).}
			\label{fig:Lpt_subsystem}
		\end{figure}
		
		\begin{algorithm}
			\caption{\texttt{Lpt_subsystem}}
			\label{alg:Lpt_subsys}
				\begin{algorithmic}
					\State \codecomm{iterate over LptDs}
					\For{$i \in I_{\tinytext{LptD}}$}
					\State \codecomm{check for transitions in insertion state}
					\If{$S^{(i)}=\textit{free}$}
					\State \codecomm{check whether to insert a new nascent LPS}
					\State Draw $x \sim \text{unif}(0,1)$
					\If{$x<1-\exp(-s_{\tinytext{LPS}}\Delta t)$}
					\State $S^{(i)}\gets\textit{embedding}$
					\State \codecomm{create new nascent LPS}
					\State Select $i^*\in \mathbb{N} \setminus I_{\tinytext{all}}$
					\State $I_{\tinytext{nasc}} \gets I_{\tinytext{nasc}}\cup \{i^*\}$
					\State Draw $\theta_{\tinytext{LPS}} \sim \text{unif}(0, 2\pi)$
					\State Set $\mathbf{r}^{(i^*)}$ using Supplementary Equation~\eqref{eq:nascent_LPS_position}
					\State \codecomm{record arrival time, inserting agent index}
					\State $t_{\tinytext{arr}}^{(i^*)}\gets t$
					\State $a_{\tinytext{ins}}^{(i^*)}\gets i$
					\EndIf
					\EndIf
					\EndFor
				\end{algorithmic}
		\end{algorithm}

		\subsubsection{Nascent agent insertion}
		\label{sec:nascent_agent_insertion}

		In this component of the model, we check each nascent agent to see if it has become fully inserted. This represents \texttt{update_nascent_agents} in Supplementary Algorithm \ref{alg:ABM_algo}.
		
		We assume that each agent type in the model requires some fixed time, depending on the type of agent, to become embedded into the membrane. For LPS, which are not assembled in the outer membrane, we assume this to be a small, fixed insertion time, $T_{\tinytext{ins,LPS}}$. For OMPs, which require templating and folding by BAM, we assume the insertion time is proportional to the radius of the OMP to be inserted. Since in this work, we only consider OmpA insertions, and not the insertion of other OMP species, we will take the OmpA insertion time to be $T_{\tinytext{ins,OmpA}}$. In this model component, we simply check how much time has elapsed since each nascent agent was inserted, and if this meets or exceeds the insertion time specific to that agent type, we promote the nascent agent to a full agent, and return its inserting agent to the \textit{free} state.
		
		We summarise this model component in Supplementary Algorithm~\ref{alg:update_nascent}

		\begin{algorithm}
			\caption{\texttt{update_nascent_agents}}
			\label{alg:update_nascent}
			\begin{algorithmic}
				\State \codecomm{iterate nascent agents}
				\For{$i \in I_{\tinytext{nasc}}$}
				\State \codecomm{check if insertion time for this agent type has elapsed}
				\If{$t - t_{\tinytext{arr}}^{(i)} \ge T_{\tinytext{ins},\gamma(i)}$}
				\State \codecomm{move nascent index to full agent indices}
				\State $I_{\tinytext{nasc}}\gets I_{\tinytext{nasc}}\setminus \{i\}$
				\State $I_{\tinytext{full}}\gets I_{\tinytext{full}}\cup\{i\}$
				\State \codecomm{return inserting agent to free state} 
				\State $S^{(a_{\tinytext{ins}}^{(i)})} \gets \textit{free}$
				\EndIf
				\EndFor
			\end{algorithmic}
		\end{algorithm}

		\subsubsection{Domain growth}
		\label{sec:domain_growth}
		
		In this component of the model, we account for growth of the spatial domain $\Omega(t)$ (and subsequent repositioning of all agents) due to the embedding of new agents. This represents \texttt{rescale_domain} in Supplementary Algorithm \ref{alg:ABM_algo}. \textbf{Note}: in the following, for ease of notation, assume that all position vectors $\mathbf{r}^{(\cdot)}(t)$, index sets $I_{\cdot}(t)$, etc. have been updated as a result of any other processes (such as BAM and Lpt subsystems) that have already taken place this time step.
		
		We account for the process of agent insertion by modelling the force acting between nascent-inserting agent pairs as a stiff spring. Both the nascent agent and its inserting agent also experience the usual attraction-repulsion forces with other nearby agents as outlined in Supplementary Section~\ref{sec:forces}. This allows the membrane to remodel as needed during the embedding process without needing to be overly prescriptive with the position of the nascent agent. Initially, the resting length of the spring is the difference in radii between the inserting and nascent agents (the same as the initial distance between the agents), and over the insertion time, the resting length of the spring increases linearly until it becomes the sum of the radii (such that the agents touch only at their boundaries). The explicit form of the nascent-inserting force and further details of implementation are given in Supplementary Section~\ref{sec:forces}. The resting length of this spring can be thought of as the ``ideal distance'' between a nascent agent and its inserting agent. The ideal distance between nascent agent $i$ and its inserting agent is given by
		\begin{equation}
			d_{\tinytext{nasc,ideal}}^{(i)}(t) = R_{\gamma(i)} - R_{\gamma(a_{\tinytext{ins}}^{(i)})} + 2\left(\frac{t-t_{\tinytext{arr}}^{ (i)}}{T_{\tinytext{ins}, \gamma(i)}}\right)R_{\gamma(i)},
			\label{eq:OmpA_ideal_distance}
		\end{equation}
		
		\noindent for $t\in[t_{\tinytext{arr}}^{(i)}, t_{\tinytext{arr}}^{(i)} + T_{\tinytext{ins}, \gamma(i)}]$.
		
		We model the increase in agent area added into the system through the insertion process by considering the exposed area of the nascent agent (that is, the portion of its area which is not overlapping with its inserting agent) when at the ideal distance. This therefore becomes a matter of circle geometry, as illustrated in Supplementary Figure~\ref{fig:nascent_exposed_area}. It is straightforward to show that at time $t\in[t_{\tinytext{arr}}^{(i)}, t_{\tinytext{arr}}^{(i)} + T_{\tinytext{ins}, \gamma(i)}]$, the exposed area of nascent agent $i$ can be given by
		\begin{equation}
			A_{\tinytext{exposed}}^{(i)}(t) = \pi R_{\gamma(i)}^2 - \frac{1}{2}\left(R_{\gamma(a_{\tinytext{ins}}^{(i)})}^2(\phi(t) - \sin(\phi(t))) + R_{\gamma(i)}^2(\psi(t) - \sin(\psi(t)))\right),
			\label{eq:OmpA_exposed_area}
		\end{equation}
		
		\noindent where $\phi(t) = \arccos(d_{\tinytext{nasc,int}}^{(i)}(t)/R_{\gamma(a_{\tinytext{ins}}^{(i)})})$, $\psi(t) = \arccos((d_{\tinytext{nasc,ideal}}^{(i)}(t)-d_{\tinytext{nasc,int}}^{(i)}(t))/R_{\gamma(i)})$, and \\$d_{\tinytext{int}}^{(i)}(t) = (R_{\gamma(a_{\tinytext{ins}}^{(i)})}^2 - R_{\gamma(i)}^2 + d_{\tinytext{nasc,ideal}}^{(i)}(t)^2)/2d_{\tinytext{nasc,ideal}}^{(i)}(t)$.
		
		\begin{figure}[h!]
			\centering
			\includegraphics[width=\textwidth, page=6, trim={3cm, 21.9cm, 3cm, 3.3cm}, clip]{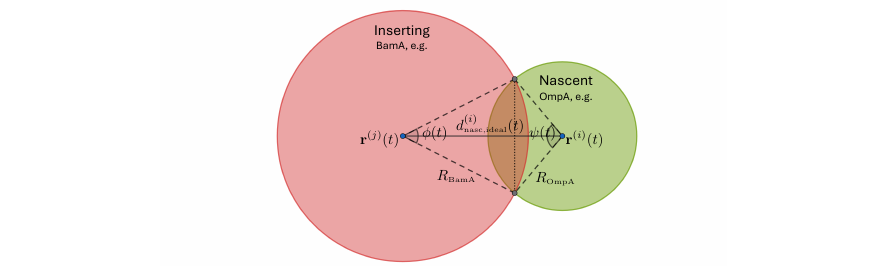}
			\caption{Calculation of the exposed area of nascent OmpA $i$ being inserted by BamA $j$ (green region not overlapping the red region). The construction is the same for nascent LPS being inserted by LptD.}
			\label{fig:nascent_exposed_area}
		\end{figure}
		
		Then, if we define
		\begin{equation}
			\tilde{A}_{\tinytext{exposed}}^{(i)}(t) = \begin{dcases}
		A_{\tinytext{exposed}}^{(i)}(t), \qquad &t\in[t_{\tinytext{arr}}^{(i)}, t_{\tinytext{arr}}^ {(i)}+T_{\tinytext{ins},\gamma(i)}],\\
		0,  &t<t_{\tinytext{arr}}^{(i)},
	    \end{dcases}
		\end{equation}
		
		\noindent the amount of exposed agent area to be added to the system at time $t$ compared to time $t -\Delta t$ can be computed as
		\begin{equation}
			\Delta A(t) = \sum_{i\in I_{\tinytext{nasc}}(t)} \left[ \tilde{A}_{\tinytext{exposed}}^{(i)}(t) - \tilde{A}_{\tinytext{exposed}}^{(i)}(t -\Delta t)\right].
		\end{equation}
		
		We allow space for nascent agents in the membrane by growing the domain as necessary. It is important that the domain is an appropriate size relative to the agent population such that the density of the agent packing emerges from the properties of the agents and is not imposed by the domain, while also not permitting holes to form in the membrane. We account for this as follows. Over any given time step, provided that the membrane is well-formed (that is, it does not contain a sufficiently large hole), we enlarge the membrane by a scaling factor such that the increase in area is equal to the total exposed agent area to be added that time step, divided by a factor which accounts for how densely agents can pack within the membrane. We have
		\begin{equation}
			\Omega(t+\Delta t) = [0, \sigma X(t)]\times[0, \sigma Y(t)] = \sigma \Omega (t)
		\end{equation}
		
		\noindent where the scaling factor for domain growth is
		\begin{equation}
			\sigma = \sqrt{1 + \frac{\Delta A(t)/\delta}{|\Omega(t)|}}
			\label{eq:scale_factor}
		\end{equation}
		
		\noindent and $\delta$ is the density factor ($\delta=0.9$, from our experience, ensures an adequate amount of area is added). The area added to the membrane is assumed to affect each agent equally, thus the position of each agent is also dilated by a factor of $\sigma$. Alternatively, if a sufficiently large hole is present in the membrane (which we define as a circle of radius $R^{\tinytext{hole}}$ that can be placed in the domain without intersecting any agent) we block domain growth until the hole closes. In this work, we consider a maximum hole around the size of an OmpA.
		
		We summarise this model component in Supplementary Algorithm \ref{alg:rescale_domain}. An illustration of the domain resizing process is given in Figure 1C in the main article.
		
		\begin{algorithm}
			\caption{\texttt{rescale_domain}}
			\label{alg:rescale_domain}
			\begin{algorithmic}
				\State \codecomm{check if the membrane contains a hole}
				\If{$\nexists \mathbf{x}^* \in \Omega~\text{s.t.} ~ d_\Omega \left( \mathbf{x}^*, \mathbf{r}^{(i)} \right) > R_{\gamma(i)} + R^{\tinytext{hole}} ~\forall i \in I_{\tinytext{all}}$}
				\State \codecomm{compute scaling factor}
				\State Compute scaling factor $\sigma$ (given by Supplementary Equation~\eqref{eq:scale_factor}).
				\State \codecomm{rescale all agent positions}
				\For{$i \in I_{\tinytext{all}}$}
				\State $\mathbf{r}^{(i)} \gets \sigma \mathbf{r}^{(i)}$
				\EndFor
				\State \codecomm{rescale domain}
				\State $\Omega \gets \sigma \Omega$
				\EndIf
			\end{algorithmic}
		\end{algorithm}

		\subsubsection{Diffusion of agents}
		\label{sec:diffusion}
		
		In this component of the model, we account for random diffusion of agents in the membrane. This represents \texttt{agent_diffusion} in Supplementary Algorithm~\ref{alg:ABM_algo}.
		
		In this work, for simplicity, we only consider diffusion of the LPS. This reflects the fact that OMPs are observed experimentally to experience much less lateral diffusion than LPS \cite{benn_et_al_OmpA_order, lithgow_et_al_surveying, horne_et_al_lipid_bilayer}. OmpA, in particular, also tethers into the peptidoglycan layer, further limiting movement \cite{benn_et_al_OmpA_order, silhavy_et_al_bacterial}. We assume that each LPS $i$ obeys Brownian motion such that during a time step $\Delta t$, it undergoes displacement $\bm{\epsilon}^{(i)}$ where
		\begin{equation}
			\bm{\epsilon}^{(i)} \sim \mathcal{N}\left(\mathbf{0}, 2\diffLPS\Delta t \mathbf{I}\right),
		\end{equation}
		
		\noindent $\mathbf{I}$ is the identity matrix and $\diffLPS$ is the diffusion coefficient.
		
		We summarise this model component in Supplementary Algorithm \ref{alg:agent_diffusion}
		
		\begin{algorithm}
			\caption{\texttt{agent_diffusion}}
			\label{alg:agent_diffusion}
			\begin{algorithmic}
				\State \codecomm{simulate diffusion of all LPS}
				\For{$i\in I_{\tinytext{LPS}}$}
				\State Draw $\bm{\epsilon}^{(i)} \sim \mathcal{N}(\mathbf{0}, 2\diffLPS\Delta t\mathbf{I})$
				\State $\mathbf{r}^{(i)}\gets \mathbf{r}^{(i)} + \bm{\epsilon}^{(i)}$
				\State \codecomm{apply periodic boundaries}
				\State $\mathbf{r}^{(i)} \gets (\mathbf{r}^{(i)}\cdot\hat{\mathbf{e}}_x  \mod X, \mathbf{r}^{(i)}\cdot\hat{\mathbf{e}}_y  \mod Y)$
				\EndFor
			\end{algorithmic}
		\end{algorithm}

		\subsubsection{Resolution of forces}
		\label{sec:forces}
		
		In this component of the model, we account for forces acting between agents in the membrane. This represents \texttt{resolve_forces} in Supplementary Algorithm \ref{alg:ABM_algo}. \textbf{Note}: in the following, for ease of notation, assume that all position vectors $\mathbf{r}^{(\cdot)}(t)$, index sets $I_{\cdot}(t)$, etc. have been updated as a result any other processes (such as BAM and Lpt subsystems diffusion, membrane resizing, etc.) that have already taken place this time step.
		
		We model interactions between all pairs of agents except nascent-inserting pairs as an attraction-repulsion force. The attraction-repulsion force experienced by agent $i$ due to agent $j$ is given by
		\begin{equation}
			\mathbf{F}_{\tinytext{attr-rep}}^{(ij)}(t) = \begin{cases}
				F_{\tinytext{rep}}^{\text{\tiny max}}\hat{\mathbf{r}}^{(ij)}(t),\qquad\qquad~  \text{for } d^{(ij)}(t) < d_{\text{\tiny min}}\\[2em]
				
				\mu_{\tinytext{rep}}^{(ij)} d_{\tinytext{nat}}^{(ij)} \log\left(\frac{ d^{(ij)}(t) - \rho d_{\tinytext{nat}}^{(ij)}}{(1- \rho)d_{\tinytext{nat}}^{(ij)}}\right)\hat{\mathbf{r}}^{(ij)}(t), \\[1em]
			    \qquad \qquad\qquad\qquad\qquad~~  \text{for } d_{\text{\tiny min}} \le d^{(ij)}(t)<d_{\tinytext{nat}}^{(ij)}\\[2em]
				
				\mu_{\tinytext{attr}}^{(ij)} d_{\tinytext{nat}}^{(ij)} \left(\frac{d^{(ij)}(t) - d_{\tinytext{nat}}^{(ij)}}{(1 - \rho) d_{\tinytext{nat}}^{(ij)}}\right)\exp\left(-k_C \frac{d^{(ij)}(t) - d_{\tinytext{nat}}^{(ij)}}{(1- \rho)d_{\tinytext{nat}}^{(ij)}}\right)\hat{\mathbf{r}}^{(ij)}(t),\\[1.5em]
				\qquad\qquad\qquad\qquad\qquad~~ \text{for } d_{\tinytext{nat}}^{(ij)}\le d^{(ij)}(t)\le d_{\tinytext{nat}}^{(ij)}+d_{\text{\tiny max}},\\[2em]
				
				\mathbf{0}, \qquad\qquad\qquad\qquad~~~~~ \text{for } d^{(ij)}(t)> d_{\tinytext{nat}}^{(ij)}+d_{\text{\tiny max},}
			\end{cases}
			\label{eq:forces}
		\end{equation}
		
		\noindent where 
		
		\begin{itemize}
			\item $d^{(ij)}(t)=d_{\Omega(t)}(\mathbf{r}^{(j)}(t),\mathbf{r}^{(i)}(t))$ is the Euclidean distance between the agents using the minimum image convention,
			\item $\mathbf{r}^{(ij)}(t)=\mathbf{v}_{\Omega(t)}(\mathbf{r}^{(j)}(t),\mathbf{r}^{(i)}(t))$ is the shortest vector from agent $j$ to agent $i$ using the minimum image convention (and $\hat{\mathbf{r}}^{(ij)}(t)$ is its corresponding unit vector),
			\item $\rho\in(0,1)$ is such that the repulsion force has a vertical asymptote at $\rho d_{\tinytext{nat}}^{(ij)}$,
			\item $F_{\tinytext{rep}}^{\text{\tiny max}}$ is the maximum repulsion force, for distances less than \\$d_{\tinytext{min}}=d_{\tinytext{nat}}^{(ij)}(\rho + (1-\rho)\exp(F_{\tinytext{rep}}^{\text{\tiny max}}/(\mu_{\text{\tiny rep}}^{(ij)} d_{\tinytext{nat}}^{(ij)})))$, 
			\item $\mu_{\text{\tiny rep}}^{(ij)}$ and $\mu_{\text{\tiny attr}}^{(ij)}$ determine the strength of the repulsion and attraction force respectively, 
			\item $k_C$ governs the decay rate of the attraction force, 
			\item $d_{\tinytext{nat}}^{(ij)}=R_{\gamma(i)}+R_{\gamma(j)}$ is the natural separation between agents $i$ and $j$ (that is, the sum of their radii),
			\item $d_{\text{\tiny max}}$ is the maximum distance between the boundaries of two agents at which the attraction-repulsion force is experienced.
	\end{itemize} 
	
	\noindent This form of the inter-agent force is similar to one proposed for multicellular dynamics by Brown \emph{et al.} \cite{brown_et_al_rigid_body}. For simplicity, we assume that the repulsion force coefficient is fixed regardless of the agent types, that is, $\mu_{\tinytext{rep}}^{(ij)}=\mu_{\tinytext{rep}}$ for all $i,j$. We also assume that the all OMP-OMP and OMP-LPS attraction force coefficients are the same regardless of the species of OMP involved. That is,
	
	\begin{equation}
		\mu_{\tinytext{attr}}^{(ij)} = \begin{dcases}
			\mu_{\tinytext{attr}}^{\tinytext{OMP-OMP}}, &\gamma(i)\in\Gamma_{\tinytext{OMP}}, \gamma(j)\in\Gamma_{\tinytext{OMP}},\\
			\mu_{\tinytext{attr}}^{\tinytext{OMP-LPS}}, &\gamma(i)\in\Gamma_{\tinytext{OMP}}, \gamma(j)=\text{LPS},\\
			\mu_{\tinytext{attr}}^{\tinytext{OMP-LPS}}, & \gamma(i)=\text{LPS}, \gamma(j)\in\Gamma_{\tinytext{OMP}},\\
			\mu_{\text{\tiny attr}}^{\tinytext{LPS-LPS}}, & \gamma(i)=\text{LPS}, \gamma(j)=\text{LPS},
		\end{dcases}
	\end{equation}
	
	\noindent where $\Gamma_{\tinytext{OMP}}=\{\text{OmpA, LptD, BamA}\}$.
	
	We plot the attraction-repulsion forces for each combination of agent types (using OmpA as a representative example of an OMP) in Supplementary Figure~\ref{fig:forces}. Parameters are as specified in Table~\ref{tab:params}. We note that, formally, force parameters should be chosen to ensure the force curve maintains a smooth derivative with respect to distance at $\mathbf{F}_{\tinytext{attr-rep}}^{(i,j)}(t)=0$ (specifically, this requires $\mu_{\tinytext{rep}}^{(i,j)}=\mu_{\tinytext{attr}}^{(i,j)}$). We have chosen not to enforce this property in this work for simplicity: since the primary purpose of the repulsion force in our model is to prevent large agent overlaps (which are unphysical anyway), it is sufficient for our purposes to consider this part of the attraction-repulsion curve to be fixed for all agent interactions.
	
	\begin{figure}[h!]
		\centering
%
%
%
		\includegraphics[width=\textwidth, page=7, trim={2cm, 16cm, 2cm, 2.5cm}, clip]{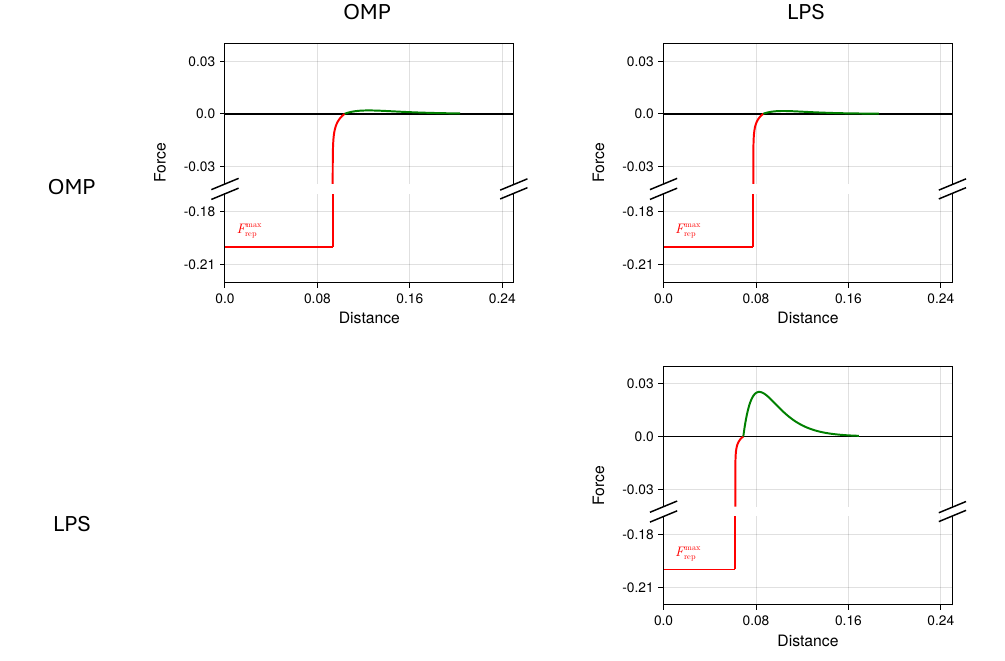}
		\caption{Attraction-repulsion forces (omitting direction) for different agent type combinations as a function of distance between agents. For the ``OMP'' cases, we show the force curves for the OmpA case an an example (different OMPs will have different natural distances etc).}
		\label{fig:forces}
	\end{figure}
	
	As mentioned in Supplementary Section~\ref{sec:domain_growth}, we model the force between nascent-inserting pairs of agents separately. The general form of the nascent-inserting force is the same as the attraction-repulsion force, but with a different parameter set and without a maximum distance at which force is experienced. The force experienced by nascent agent $i$ due to its inserting agent $j$ is given by
	
	\begin{equation}
		\mathbf{F}_{\tinytext{nasc-ins}}^{(ij)}(t) = \begin{cases}
			F_{\tinytext{rep}}^{\text{\tiny max}}\hat{\mathbf{r}}^{(ij)}(t),\qquad\qquad\qquad  \text{for } d^{(ij)}(t) <\rho d_{\tinytext{nasc,ideal}}^{(i)}(t), \\[2em]
			
			\mu_{\text{\tiny rep,nasc-ins}} d_{\tinytext{nasc,ideal}}^{(i)}(t) \log\left(\frac{ d^{(ij)}(t) - \rho d_{\tinytext{nasc,ideal}}^{(i)}(t)}{(1- \rho)d_{\tinytext{nasc,ideal}}^{(i)}(t)}\right)\hat{\mathbf{r}}^{(ij)}(t), \\[1.5em]
			\qquad \qquad\qquad\qquad\qquad~~~~~  \text{for } \rho d_{\tinytext{nasc,ideal}}^{(i)}(t) \le d^{(ij)}(t)< d_{\tinytext{nasc,ideal}}^{(i)}(t),\\[2em]
			
			\mu_{\text{\tiny attr,nasc-ins}} d_{\tinytext{nasc,ideal}}^{(i)}(t) \left(\frac{d^{(ij)}(t) - d_{\tinytext{nasc,ideal}}^{(i)}(t)}{(1 - \rho) d_{\tinytext{nasc,ideal}}^{(i)}(t)}\right)\exp\left(-k_C \frac{d^{(ij)}(t) -  d_{\tinytext{nasc,ideal}}^{(i)}(t)}{(1- \rho)d_{\tinytext{nasc,ideal}}^{(i)}(t)}\right)\hat{\mathbf{r}}^{(ij)}(t),\\[1.5em]
			\qquad\qquad\qquad\qquad\qquad~~~~~ \text{for } d_{\tinytext{nasc,ideal}}^{(i)}(t)\le d^{(ij)}(t).
		\end{cases}
		\label{eq:nascent-inserting_force}
	\end{equation}
	
	\noindent The force experienced by the inserting agent, $j$, as a result of its nascent agent, $i$, is simply the negative:
	\begin{equation}
		\mathbf{F}_{\tinytext{nasc-ins}}^{(ji)}(t) = - \mathbf{F}_{\tinytext{nasc-ins}}^{(ij)}(t).
	\end{equation}
	
	\noindent Then for any two agents $i$ and $j$, the force between them is given by
	\begin{equation}
		\mathbf{F}^{(ij)}(t) = \begin{dcases}
			\mathbf{F}_{\tinytext{nasc-ins}}^{(ij)}(t), &i\in I_{\tinytext{nasc}}(t), j=a_{\tinytext{ins}}(i),\\
			- \mathbf{F}_{\tinytext{nasc-ins}}^{(ji)}(t), &j\in I_{\tinytext{nasc}}(t), i=a_{\tinytext{ins}}(j),\\
			\mathbf{F}_{\tinytext{attr-rep}}^{(ij)}, &\text{otherwise}.
		\end{dcases}
	\end{equation}
		
		The position of each agent $i$ under the action of the inter-agent forces is given by overdamped Newtonian mechanics
		\begin{equation}
			\eta_{\gamma(i)} \ddt{}\mathbf{r}^{(i)}(t) = \sum_{j\in I_{\tinytext{all}}(t) \setminus \{i\}} \mathbf{F}^{(ij)}(t)
		\end{equation}
		
		\noindent where $\eta_{\gamma^*} = \eta R_{\gamma^*}$ is the drag coefficient associated with agents of type $\gamma^*$, assuming (in line with Atwell \emph{et al.} Development 2015) that the drag coefficient of each agent is proportional to its radius. We deduce the following explicit update scheme for the position of each agent $i$ over the time interval $[t, t+\Delta t]$:
		\begin{equation}
			\mathbf{r}^{(i)}(t+\Delta t) = \mathbf{r}^{(i)}(t) + \frac{ \Delta t}{\eta R_{\gamma(i)}}\sum_{i \in I_{\tinytext{all}}(t) \setminus \{i\}} \mathbf{F}^{(ij)}(t).
		\end{equation}
		
		In practice, this model component is implemented as described in Supplementary Algorithm~\ref{alg:resolve_forces}. 
		
		\begin{algorithm}
			\caption{\texttt{resolve_forces}}
			\label{alg:resolve_forces}
			\begin{algorithmic}
				\State \codecomm{iterate over agents}
				\For{$i\in I_{\tinytext{all}}$}
				\State \codecomm{tally all forces}
				\State $\mathbf{F}^{(i)}\gets \mathbf{0}$
				\For{$j \in I_{\tinytext{all}} \setminus \{i\}$}
				\If{$i \in I_{\tinytext{nasc}} \textbf{ and } j=a_{\tinytext{ins}}^{(i)}$}
				\State \codecomm{agent is nascent, add force from inserting agent}
				\State Compute $\mathbf{F}_{\tinytext{nasc-ins}}^{(ij)}$ (given by Supplementary Equation \eqref{eq:nascent-inserting_force})
				\State $\mathbf{F}^{(i)}\gets \mathbf{F}^{(i)} + \mathbf{F}_{\tinytext{nasc-ins}}^{(ij)}$
				\ElsIf{$j \in I_{\tinytext{nasc}} \textbf{ and } i=a_{\tinytext{ins}}^{(j)}$}
				\State \codecomm{agent is inserting a nascent agent, add force from nascent agent}
				\State Compute $\mathbf{F}_{\tinytext{nasc-ins}}^{(ji)}$ (given by Supplementary Equation \eqref{eq:nascent-inserting_force})
				\State $\mathbf{F}^{(i)}\gets \mathbf{F}^{(i)} - \mathbf{F}_{\tinytext{nasc-ins}}^{(ji)}$
				\Else
				\State \codecomm{not nascent-inserting pair, compute attraction-repulsion force}
				\State Compute $\mathbf{F}_{\tinytext{attr-rep}}^{(ij)}$ (given by Supplementary Equation \eqref{eq:forces})
				\State $\mathbf{F}^{(i)}\gets \mathbf{F}^{(i)} + \mathbf{F}_{\tinytext{attr-rep}}^{(ij)}$
				\EndIf
				\EndFor
				\State \codecomm{update position}
				\State $\mathbf{r}^{(i)} \gets \mathbf{r}^{(i)}(t) + \Delta t/(\eta R_{\gamma(i)})\mathbf{F}^{(i)}$
				\State \codecomm{apply periodic boundaries}
				\State $\mathbf{r}^{(i)} \gets (\mathbf{r}^{(i)}\cdot\hat{\mathbf{e}}_x  \mod X, \mathbf{r}^{(i)}\cdot\hat{\mathbf{e}}_y  \mod Y)$
				\EndFor
			\end{algorithmic}
		\end{algorithm}

		\clearpage
		
		\section{Parameters}
		\label{sec:parameters}
		
		Parameters are divided into the following groups: 
		\begin{itemize}
			\item \colorbox{teal!30}{Biological parameters} --- Drawn from the Research Collaboratory for Structural Bioinformatics (RCSB) Protein Data Bank.
			\item \colorbox{yellow!40}{Insertion parameters} --- Selected to ensure sufficient time for the system to relax between agent insertions.
			\item \colorbox{purple!30}{Force parameters} --- Attraction force and diffusion coefficients varied in the manuscript; other values selected to ensure reasonable dynamics, especially in preventing overlaps. Nascent-inserting force parameters chosen to be relatively stiff compared to attraction-repulsion forces.
			\item \colorbox{brown!30}{Equilibration parameters} --- Selected through trial and error to prevent holes and allow membrane systems to properly equilibrate.
			\item \colorbox{lightgray}{Numerical parameters} --- Selected through trial and error to prevent numerical errors for the other chosen parameters (especially force parameters).
		\end{itemize}
		
		In the following we take $L$ to be the spatial unit (around 174\r{A}) and $T$ to be the time unit (future parameter estimation against experimental data is necessary to fully quantify these units). Here, we write $N^* = \text{Da} L T^{-2}$ as the force unit.
		
		\begingroup
		\renewcommand{\arraystretch}{1.5}
		\begin{center}
		\begin{longtable}{|L{2cm} |L{2.5cm} |L{7cm} |L{2.5cm}|}
			\hline
			
			\rowcolor{lightgray} \textbf{Symbol}&\textbf{Value}&\textbf{Description}&\textbf{Source}\\
			
			\hline
			\hline
			
			\rowcolor{teal!30}
			$R_{\tinytext{OmpA}}$&
			0.0518$L$&
			OmpA radius&
			\cite{pautsch_OmpA_PDB}\\
			\hline
			
			\rowcolor{teal!30}
			$R_{\tinytext{BamA}}$&
			0.1725$L$&
			BamA radius&
			\cite{albrecht_et_al_BamA_structure}\\
			\hline
			
			\rowcolor{teal!30}
			$R_{\tinytext{LptD}}$&
			0.259$L$&
			LptD radius&
			\cite{dong_et_al_structural}\\
			\hline
			
			\rowcolor{teal!30}
			$R_{\tinytext{LPS}}$&
			0.0345$L$&
			LPS radius&
			\cite{dong_et_al_structural, lebrun_et_al_structural}\\
			\hline

			\rowcolor{yellow!40}
			$s_{\tinytext{poly}}$&
			0.5 polypeptides $T^{-1}$&
			Polypeptide arrival rate&
			Assumed\\
			\hline
			
			\rowcolor{yellow!40}
			$\beta$&
			0.05 $L^2$ polypeptides$^{-1}$ $T^{-1}$&
			Polypeptide to BAM binding rate&
			Assumed\\
			\hline
			
			\rowcolor{yellow!40}
			$T_{\tinytext{ins,OmpA}}$&
			2.072 $T$&
			OmpA insertion time&
			Assumed$^\dagger$\\
			\hline
			
			\rowcolor{yellow!40}
			$T_{\tinytext{ins,LPS}}$&
			1.0 $T$&
			LPS insertion time&
			Assumed$^\dagger$\\
			\hline
			
			\rowcolor{yellow!40}
			$R^{\tinytext{sens}}$&
			0.1 $L$&
			LPS sensing radius for BAM insertion&
			Assumed\\
			\hline

			\rowcolor{purple!30}
			$\mu_{\tinytext{attr}}^{\tinytext{OMP-OMP}}$&
			0.025 $N^* L^{-1}$ [default]&
			OMP-OMP attraction force coefficient&
			Varied in manuscript\\
			\hline
			
			\rowcolor{purple!30}
			$\mu_{\tinytext{attr}}^{\tinytext{OMP-LPS}}$&
			0.025 $N^* L^{-1}$ [default]&
			OMP-LPS attraction force coefficient&
			Varied in manuscript\\
			\hline
			
			\rowcolor{purple!30}
			$\mu_{\tinytext{attr}}^{\tinytext{LPS-LPS}}$&
			0.1 $N^* L^{-1}$ [default]&
			LPS-LPS attraction force coefficient&
			Varied in manuscript\\
			\hline
			
			\rowcolor{purple!30}
			$\mu_{\tinytext{rep}}$&
			0.05 $N^* L^{-1}$&
			Agent-agent repulsion force coefficient&
			Assumed$^\dagger$\\
			\hline
			
			\rowcolor{purple!30}
			$k_C$&
			0.5&
			Attraction force decay coefficient&
			Assumed\\
			\hline
			
			\rowcolor{purple!30}
			$\rho$&
			0.9&
			Location of vertical asymptote for inter-agent repulsion force (proportion of inter-agent distance)&
			Assumed$^\dagger$\\
			\hline
			
			\rowcolor{purple!30}
			$F_{\tinytext{rep}}^{\tinytext{max}}$&
			-0.2 $N^*$&
			Maximum repulsion force&
			Assumed$^\dagger$\\
			\hline
			
			\rowcolor{purple!30}
			$d_{\tinytext{max}}$&
			0.1 $L$&
			Maximum distance between agent boundaries at which the attraction-repulsion force is experienced&
			Assumed$^\dagger$\\
			\hline
			
			\rowcolor{purple!30}
			$\mu_{\tinytext{attr,nasc-ins}}$&
			0.5 $N^* L^{-1}$&
			Nascent-inserting attraction force coefficient&
			Assumed$^\dagger$\\
			\hline
			
			\rowcolor{purple!30}
			$\mu_{\tinytext{rep,nasc-ins}}$&
			0.5 $N^* L^{-1}$&
			Nascent-inserting repulsion force coefficient&
			Assumed$^\dagger$\\
			\hline
			
			\rowcolor{purple!30}
			$\eta$&
			1.0 $L^{-1}$&
			Drag coefficient (proportional to agent radius)&
			Assumed$^\ddagger$\\
			\hline
			
			\rowcolor{purple!30}
			$D$&
			0.0 $L^2 T^{-1}$ [default]&
			LPS diffusion coefficient&
			Varied in manuscript\\
			\hline

			\rowcolor{brown!30}
			$t_{\tinytext{eqm}}$&
			10.0 $T$&
			Initial system equilibration time&
			Tuned\\
			\hline
			
			\rowcolor{brown!30}
			$t_{\tinytext{short eqm}}$&
			2.0 $T$&
			Re-equilibration time during domain shrinkage&
			Tuned\\
			\hline
			
			\rowcolor{brown!30}
			$\sigma_{\tinytext{shrink}}$&
			3.0 $T$&
			Scaling factor for domain shrinkage&
			Tuned\\
			\hline
			
			\rowcolor{brown!30}
			$R^{\tinytext{hole}}$&
			0.05 $L$&
			Maximum allowable membrane hole radius&
			Tuned\\
			\hline

			\rowcolor{lightgray}
			$\delta$&
			0.9&
			Density factor&
			Tuned\\
			\hline
			
			\rowcolor{lightgray}
			$\Delta t$&
			0.0001 $T$&
			Simulation time step&
			Tuned\\
			\hline

			\hline
			
			\caption{Parameter values used in the main manuscript. $^\dagger$These parameter values are not necessarily meant to convey any particular physical meaning, they are simply selected to ensure numerical stability, prevent agent overlaps etc. $^\ddagger$The drag coefficient can be absorbed into the other force parameters, hence is here assumed to be 1.}
			\label{tab:params}
		\end{longtable}
		\end{center}
		\endgroup

		\clearpage
		\bibliographystyle{apalike}